\documentclass[twoside,twocolumn,9pt]{article}
\usepackage{extsizes}
\usepackage[super,sort&compress,comma]{natbib} 
\usepackage[version=3]{mhchem}
\usepackage[left=1.5cm, right=1.5cm, top=1.785cm, bottom=2.0cm]{geometry}
\usepackage{balance}
\usepackage{times,mathptmx}
\usepackage{sectsty}
\usepackage{graphicx} 
\usepackage{lastpage}
\usepackage[format=plain,justification=justified,singlelinecheck=false,font={stretch=1.125,small,sf},labelfont=bf,labelsep=space]{caption}
\usepackage{float}
\usepackage{fancyhdr}
\usepackage{fnpos}
\usepackage[english]{babel}
\addto{\captionsenglish}{%
  
}
\usepackage{array}
\usepackage{droidsans}
\usepackage{charter}
\usepackage[utf8]{inputenc}

\usepackage[T1]{fontenc}
\usepackage[usenames,dvipsnames]{xcolor}
\usepackage{setspace}
\usepackage[compact]{titlesec}
\usepackage{hyperref}
\usepackage{siunitx}

\usepackage{epstopdf}

\definecolor{cream}{RGB}{222,217,201}

\begin{document}

\pagestyle{fancy}
\thispagestyle{plain}
\fancypagestyle{plain}{

\fancyhead[C]{\includegraphics[width=18.5cm]{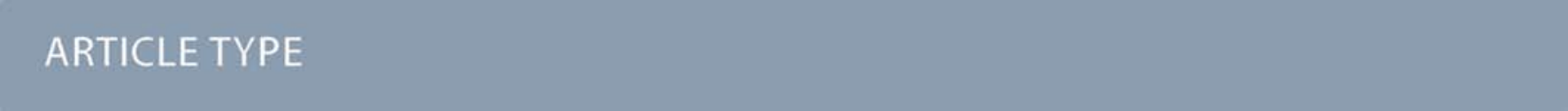}}
\fancyhead[L]{\hspace{0cm}\vspace{1.5cm}\includegraphics[height=30pt]{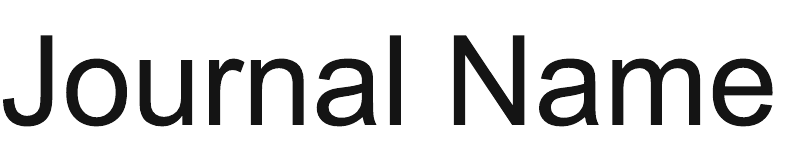}}
\fancyhead[R]{\hspace{0cm}\vspace{1.7cm}\includegraphics[height=55pt]{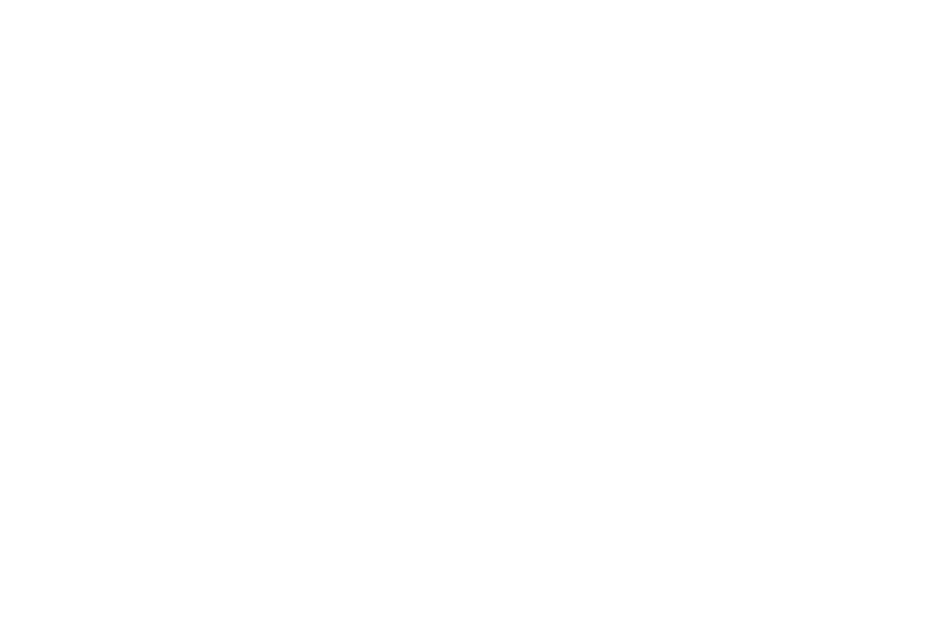}}
\renewcommand{\headrulewidth}{0pt}
}

\makeFNbottom
\makeatletter
\renewcommand\LARGE{\@setfontsize\LARGE{15pt}{17}}
\renewcommand\Large{\@setfontsize\Large{12pt}{14}}
\renewcommand\large{\@setfontsize\large{10pt}{12}}
\renewcommand\footnotesize{\@setfontsize\footnotesize{7pt}{10}}
\makeatother

\renewcommand{\thefootnote}{\fnsymbol{footnote}}
\renewcommand\footnoterule{\vspace*{1pt}%
\color{cream}\hrule width 3.5in height 0.4pt \color{black}\vspace*{5pt}} 
\setcounter{secnumdepth}{5}

\makeatletter 
\renewcommand\@biblabel[1]{#1}            
\renewcommand\@makefntext[1]%
{\noindent\makebox[0pt][r]{\@thefnmark\,}#1}
\makeatother 
\renewcommand{\figurename}{\small{Fig.}~}
\sectionfont{\sffamily\Large}
\subsectionfont{\normalsize}
\subsubsectionfont{\bf}
\setstretch{1.125} 
\setlength{\skip\footins}{0.8cm}
\setlength{\footnotesep}{0.25cm}
\setlength{\jot}{10pt}
\titlespacing*{\section}{0pt}{4pt}{4pt}
\titlespacing*{\subsection}{0pt}{15pt}{1pt}

\fancyfoot{}
\fancyfoot[LO,RE]{\vspace{-7.1pt}\includegraphics[height=9pt]{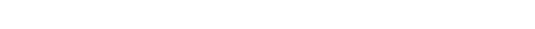}}
\fancyfoot[CO]{\vspace{-7.1pt}\hspace{13.2cm}\includegraphics{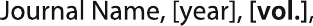}}
\fancyfoot[CE]{\vspace{-7.2pt}\hspace{-14.2cm}\includegraphics{head_foot/RF}}
\fancyfoot[RO]{\footnotesize{\sffamily{1--\pageref{LastPage} ~\textbar  \hspace{2pt}\thepage}}}
\fancyfoot[LE]{\footnotesize{\sffamily{\thepage~\textbar\hspace{3.45cm} 1--\pageref{LastPage}}}}
\fancyhead{}
\renewcommand{\headrulewidth}{0pt} 
\renewcommand{\footrulewidth}{0pt}
\setlength{\arrayrulewidth}{1pt}
\setlength{\columnsep}{6.5mm}
\setlength\bibsep{1pt}

\makeatletter 
\newlength{\figrulesep} 
\setlength{\figrulesep}{0.5\textfloatsep} 

\newcommand{\topfigrule}{\vspace*{-1pt}%
\noindent{\color{cream}\rule[-\figrulesep]{\columnwidth}{1.5pt}} }

\newcommand{\botfigrule}{\vspace*{-2pt}%
\noindent{\color{cream}\rule[\figrulesep]{\columnwidth}{1.5pt}} }

\newcommand{\dblfigrule}{\vspace*{-1pt}%
\noindent{\color{cream}\rule[-\figrulesep]{\textwidth}{1.5pt}} }

\makeatother

\twocolumn[
  \begin{@twocolumnfalse}
\vspace{3cm}
\sffamily
\begin{tabular}{m{4.5cm} p{13.5cm} }

\includegraphics{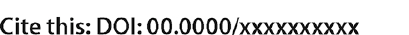} & \noindent\LARGE{\textbf{Electronic transport properties of \ce{MoS2} nanoribbons embedded on butadiene solvent}} \\
\vspace{0.3cm} & \vspace{0.3cm} \\

 & \noindent\large{Armando Pezo,$^{\ast}$\textit{$^{a,b}$} Matheus P. Lima,\textit{$^{c}$} Marcio Costa,\textit{$^{a}$} and Adalberto Fazzio$^{\ast}$\textit{$^{a}$}} \\

\includegraphics{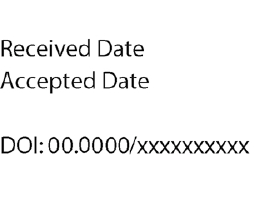} & \noindent\normalsize{
Transition metal dichalcogenides (TMDCs) are promising materials for
applications in nanoelectronics and correlated fields, where their metallic edge
states play a fundamental role in the electronic transport. In this work, we
investigate the transport properties of \ce{MoS2} zigzag nanoribbons under a
butadiene (\ce{C4H6}) atmosphere, 
as this compound has been used to obtain \ce{MoS2} flakes by exfoliation. 
We use the density functional theory combined to non-equilibrium Green's
functions techniques, in a methodology contemplating disorder and different
coverages. Our results indicate a strong modulation of the TMDC electronic transport
properties driven by butadiene molecules anchored at their edges,
producing the suppression of currents due to a backscattering process.
Our results indicate a high sensibility of the TMDC edge states.  Thus, the mechanisms 
used to reduce the dimensionality of \ce{MoS2} considerably modify its transport properties. 
} \\

\end{tabular}

 \end{@twocolumnfalse} \vspace{0.6cm}

  ]

\renewcommand*\rmdefault{bch}\normalfont\upshape
\rmfamily
\section*{}
\vspace{-1cm}


\footnotetext{\textit{$^{a}$~Brazilian Nanotechnology National Laboratory (LNNano,CNPEM), Rua Giusepe M\'aximo Scalfaro 10000, Campinas, SP 13083-970, Brazil; E-mail: armando.pezo@ufabc.edu.br, adalberto.fazzio@lnnano.cnpem.br}}
\footnotetext{\textit{$^{b}$~CCNH - Center for Natural Sciences and Humanities, Federal University of ABC, Santo Andr{\'e}, SP, Brazil}}
\footnotetext{\textit{$^{c}$~Departamento de F\'{\i}sica, Universidade Federal de S\~{a}o Carlos, 13565-905 S\~{a}o Carlos, S\~ao Paulo, Brasil}}




\section{Introduction}

Graphene single layers were isolated a decade ago~\cite{Novoselov666}
challenging conjectures about the potential existence of two dimensional
(2D) materials. This event initiated the 2D materials era\cite{2Dera}. 
However, even with its remarkable properties, such as high charge carrier 
mobility\cite{Novoselov2005}, mechanical stability\cite{Kim2009}, thermal conductivity\cite{doi:10.1021/nl0731872}, and interesting optical
properties\cite{Falkovsky_2008}, its practical use is not fully materialized. The existence of a 
gapless electronic spectrum near the Fermi level together with a difficulty 
for obtaining clean edges is a drawback for some applications. 
The so-called Klein tunneling, unconventional Hall
effect, and anti-localization behavior are related to this feature~\cite{RevModPhys.81.109}.
%

Following graphene, a profusion of 2D materials has been proposed and synthesized.
Amongst them, the transition metal dichalcogenides (TMDC) occupies a special
place. Its chemical formula \ce{\textit{MX}2}  (\ce{\textit{M}}= transition metal, and \ce{\textit{X}}=\ce{S}, \ce{Se}, or \ce{Te}) demonstrates the possibility of numerous distinct compositions with different
characteristics~\cite{Han2015,2053-1583-2-2-022001,Ataca-1}. 
Furthermore, the crystal symmetry (2H, 1T or 1T$^\prime$) is another tuning
parameter allowing to access semiconductor, metallic or topological insulator
phases for some particular compositions\cite{Qian1344,chem2}. The study of other types of 2d layered materials was achieved by looking at the edges terminations and even exploring lateral heterostructures leading to interesting results related to their I/V characteristics predicting negative differential resistance (NDR) features. 
\cite{An_2018, 8443152, AN201743, C7CP02201H, C6TC04327E,C8CP04272A}.
The broad variety of properties found in TMDCs together with the possibility of
its control by several mechanisms opens the possibility of some promising technological applications, such as hydrodesulfurization catalyst\cite{doi:10.1021/acsnano.5b00501}, thermoelectric devices\cite{C7TC01088E}, photovoltaic cells\cite{Akama2017}, and nanotribology~\cite{C2JM15906F,Ataca-2}, just to mention a few. It's worth to mention that it's been argued that single(few) layer(s) \ce{MoS2} could be used for logical electronic devices available to work at room temperature, being these thinner than the actual silicon-based films used in technological applications. The smaller dielectric constant with respect to that one of silicon is another advantage, leading to a less power consumption in transistors made of them. Measurements made on samples of this material at a bias of \SI{500}{\milli\volt} gave a maximal on-current of \SI{22}{\micro\ampere} (or \SI{4.6}{\micro\ampere/\micro\metre}) at room temperature with a large $I_{on}/I_{off}$ ratio higher than \num{e6} for a \SI{\pm 4}{\volt} gate voltage range\cite{doi:10.1021/nn303513c,doi:10.1021/nl2018178}. 


However, the integration of TMDCs in future technologies in many cases requests the reduction of its dimensionality. In this regard, the electronic properties of \ce{MoS2} nanoribbons and nanoflakes are strongly constrained by their edge termination, whereas metallic and ferromagnetic edge states arise for the zigzag border. On the other hand, the armchair termination shows semiconducting and non-magnetic behavior, although it might transform into a metallic and magnetic under edge H-passivation~\cite{0022-3727-49-24-245304,C2JM15906F}.

There are several methods to reduce the dimensionality of \ce{MoS2}, ranging form chemical to mechanical routes~\cite{chem1,chem2}. Recently, Gon\c{c}alves and coworkers used an assisted mechanical exfoliation to produce large \ce{MoS2} flakes\cite{doi:10.1002/chem.201502303}. They used a butadiene solution followed by an ultrasonication process, after which it is found a colloidal stability of \ce{MoS2} in the non-polar solvent, related to the chemical bonding between the molecules and atoms along the sample edges~\cite{doi:10.1002/chem.201502303}. 
In this sense, the edge generation in \ce{MoS2} through techniques assisted by butadiene molecules probably result in systems with those molecules adsorbed in \ce{MoS2} bulk and edge sites. Thus, having in mind eventual future nanotechnology applications, it is essential  to know  some fundamental characteristics such as (i) the most likely adsorption sites and the corresponding stable geometries;  (ii) Details of the adsorption mechanism (physisorption/chemisorption, binding energy, etc.); (iii) and the role of butadiene molecules in the electronic transport of \ce{MoS2}. The main objective of our work is to shed light in those questions. Molecules and atoms absorbed in different 2d-materials studies provide excellent candidates for gas sensor and H-storage devices as is the case of phosphorene and \ce{MoS2} itself  \cite{Yue2013, doi:10.1021/jp2097026, doi:10.1021/jp510899m, doi:10.1021/jp510863p, C8CP05093G, doi:10.1021/acs.jpcc.7b09552} . The binding energies presented in previous reports for absorbed atoms and molecules in \ce{MoS2} are comparable with ours results. Furthermore, Mo-terminated edges have been demonstrated to be more active, {\it i.e} more energetically favorable, when compared to the S-termination, which is in line with our findings\cite{doi:10.1021/jp807705y, doi:10.1021/jp1115146}.


In this work, we investigate the adsorption of butadiene molecules
in \ce{MoS2} nanoribbons and its influence in the electronic transport 
properties.  We employ nanoribbons with zigzag edge termination (ZZNRs). 
Our results were obtained with a methodology based on {\it ab initio} 
density functional theory (DFT)  calculations, combined with the 
recursive non-equilibrium Green's function techniques,  taking into 
account the effects of disorder and different coverages. Such a 
methodology allows a full description of the 
transport properties of large-scale devices. We show that butadiene 
molecules adsorbed on the central region of the nanoribbons do not change 
the electric current, whereas adsorption at the edge sites (which are 
the energetically preferable ones) suppresses the charge flowing 
through the metallic edge states due to a backscattering process. 
Our findings demonstrate that the low energy edge transport channels 
of TMDCs are highly susceptible to the adsorption of molecules. Thus, 
the route used to reduce the dimensionality of TMDCs will have great influence 
in the transport properties.


\section{Computational details}

An accurate atomistic simulation of \ce{MoS2} with edges and
adsorbed molecules requires a methodology with a great 
prediction capability, which is the case of the 
first-principle calculations based on DFT \cite{DFT1,review}
we used. The spin-polarized local-density-approximation (LDA)
\cite{DFT1} is used for the Exchange-correlation functional. 
We perform all geometry optimizations with the PAW 
approach, as implemented in the vienna ab-initio simulation 
package (VASP)\cite{vasp1,vasp2}. In these calculations, 
we employ \SI{400}{\electronvolt} for the 
plane-wave expansion cutoff, a force criterion of \SI{0.01}{\electronvolt/\angstrom}, 
and sample the reciprocal space with 10 $\vec{k}$-points along the 
periodic direction of our nanoribbons.

For the electronic transport calculations, we employ a methodology that couples NEGF techniques to DFT calculations. This methodology considers a central scattering region sandwiched between the left and right electrodes. We calculated the  electric current $I$ in the ballistic approach by the Landauer formula:\cite{landauer1} 
\begin{multline}
I=\frac{e}{\hbar\pi}\int_{-\infty}^{+\infty} T(E)[ f_{FD}(E-\mu_L)-f_{FD}(E-\mu_R) ]dE
\end{multline}
$e$ is the elementary charge, $\hbar$ the reduced Planck's constant, 
$f_{FD}$ the Fermi-Dirac distribution, $E$ the energy of the charge
carrier passing through the system, and $\mu_{L(R)}$ the chemical potential of the left (right) electrode.


$T(E)$ is the transmission probability function, which we evaluate from
DFT calculations using the Kohn-Sham Hamiltonian, as 
described in Refs. \cite{NOVAES2006} and \cite{PhysRevB.65.165401}.
Once sparse matrices are necessary to apply this methodology, 
it is mandatory to adopt a localized basis set to expand the Kohn-Sham
orbitals. This methodology was successfully used to investigate the
transport properties of graphene-like
materials\cite{PhysRevB.84.245411,8387867}. Furthermore, to take into
account the effects of disorder 
in large scale systems, the methodology presented in Refs.
\cite{recursive1} and \cite{recursive2} are employed. 
The Hamiltonian used to generate the $T(E)$ in a large scale system 
can be written by:

\begin{eqnarray}
H=\left(\begin{array}{ccccc}
H_L & H_C & \text{\huge0} & \hdots & \text{\huge0} \\
H_C^\dagger & H_i & H_C & \text{\huge0} &\vdots \\
 \text{\huge0} & \text{\huge0} & \ddots &\vdots &\text{\huge0} \\
 \vdots & \vdots & H_C^\dagger  &  H_j & H_C \\
\text{\huge0} & \hdots & \text{\huge0} & H_C^\dagger & H_R \\
\end{array}\right)
\end{eqnarray}

\begin{figure}
    \centering
    \includegraphics[width=8.5cm]{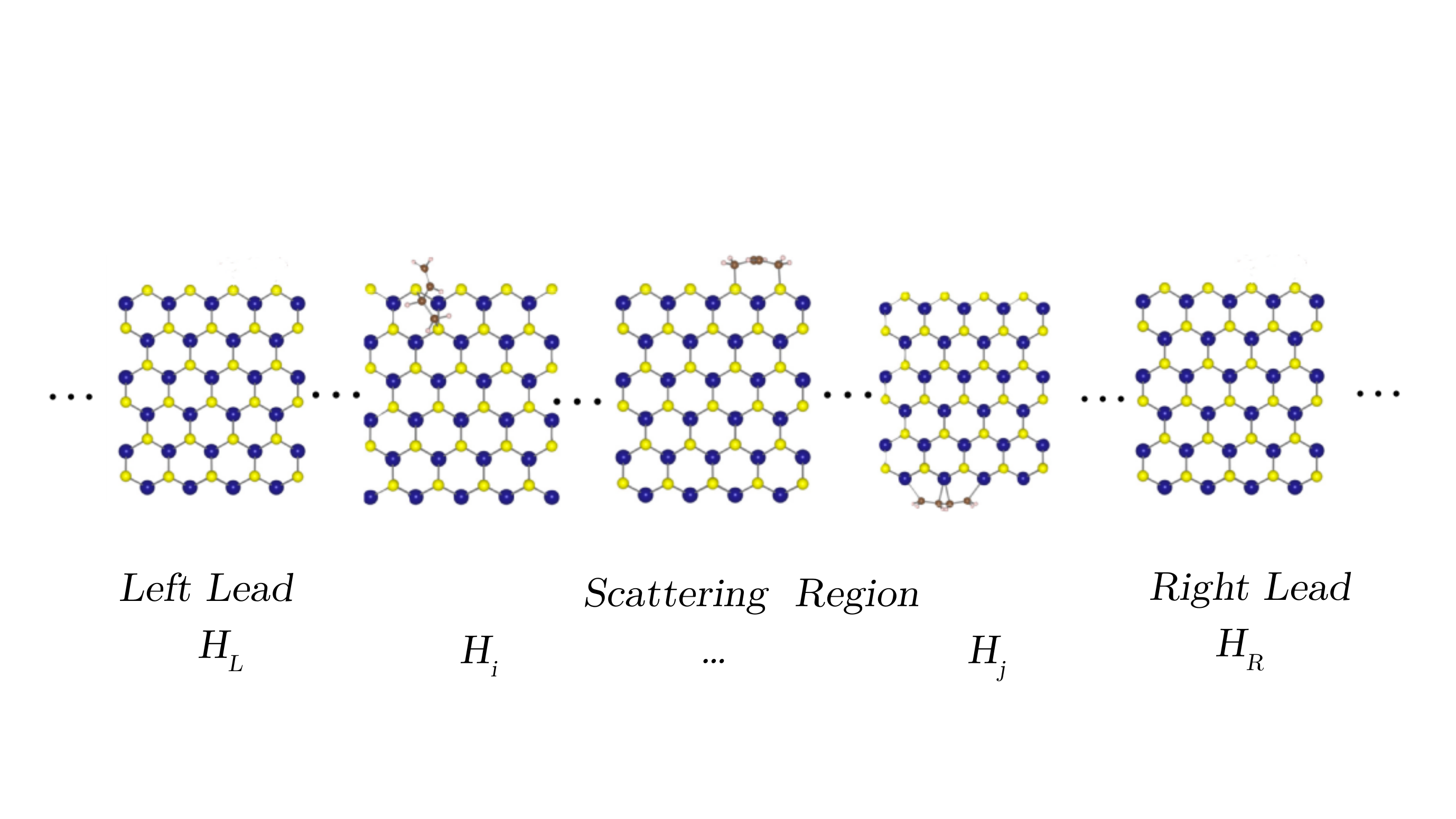}
    \caption{Schematic representation of the building blocks used to construct the large scale hamiltonian. }
    \label{fig:EsqDis}
\end{figure}

We show the geometric interpretation of the equation (1) in Fig.  
\ref{fig:EsqDis}. $H_i$ and $H_j$ are matrix elements of 
parts of the scattering region and each one represents either 
pristine sectors, or fragments  of the system containing defects, 
adsorbed molecules, or any other perturbation. In our work, we 
constructed the central region by either pristine fragments or 
nanoribbon sectors with butadiene adsorbed in various adsorption sites.
Building blocks randomly positioned along the transport direction 
are used to construct the scattering region. $H_C$ is the coupling 
matrix, and we guarantee the same $H_C$ between all building 
blocks including buffer layers in the geometries. $H_{L(R)}$ are 
the Hamiltonians of the left and right electrodes. 

We calculate the matrix elements of each building block $H_i$ and 
the coupling matrix $H_C$ via standard DFT calculations. For this
purpose, we employ a localized basis set approach as implemented in 
the SIESTA code. We used a cutoff energy of \SI{250}{Ry} to defined the 
mesh size in the real space. 

Nanoribbons with 6 (and 12 in few cases) zigzag lines from one edge to the other are used 
to generate our results. Buffer layers of two unit cells are
used to ensure the same coupling term $H_C$ between all building
blocks. The pristine blocks and the electrodes have 
two unit cells along the transport direction, whereas the blocks 
with adsorbed molecules have seven unit cells. We use a vacuum 
of \SI{20}{\angstrom} to separate the periodic images of the nanoribbons.


\section {Results}

To present our results, we first discuss the structural and energetic properties, 
in the section \ref{sec:ener}. Subsequently, the transport properties including the disorder 
effects are shown, in section \ref{sec:trans}.

\subsection{Structural and energetic properties\label{sec:ener}}

\subsubsection{Pristine nanoribbon.}
\begin{figure}[t]
\centering
\includegraphics[scale=0.3]{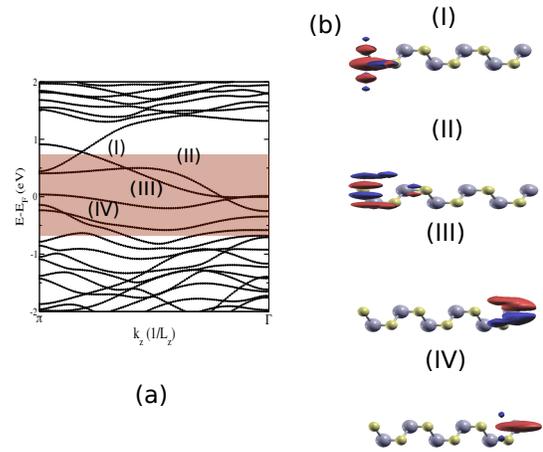}
\caption{\ce{MoS2} NR's  band structure(a) and (b) projected Wannier states on the pristine structure. The plot shows how are localized along the edges those states leaving in an energy window containing the Fermi level (around \SI{\pm 0.5}{\electronvolt}). The two different colors represent the positive(red) and negative(blue) real amplitudes for each Wannier orbital.}
\label{fig:test}
\end{figure}

First of all, we focus on the band structure of a TMDC nanoribbon with six zigzag lines, 
depicted in the left panel of Fig. \ref{fig:test} (a). Its metallic behavior is clear from the bands crossing the Fermi level (which is set at zero). Starting from the previously converged density matrix, we made a Wannier interpolation \cite{MOSTOFI20142309}, using an increased 1x30x1 k-point grid with 30 bands selected such that this set reproduce the band structure for the states surrounding the Fermi level, in particular, the states with pronounced dispersion which will be shown to live at the edges. In this manner, the required construction of the overlap and projections matrices was done by internal routines developed by SIESTA using for this purpose previously selected \ce{Mo} and \ce{S} orbitals as an initial basis set. In order to obtain an accurate description in terms of these Wannier functions, this representation must converge to that of one that represents a maximally localized set. Once the above procedure is performed we arrive to the localized orbitals depicted at Fig\ref{fig:test} in (b), where the enumerated bands are represented along the edges of the nanoribbon. It is worth to note the edge localization 
of the metallic states, indicating that the edges of the ribbons are the 
most active regions for the charge transport.  
Furthermore, the zigzag nanoribbons edges are not symmetrical. 
One side has a \ce{Mo}-rich termination, whereas the other side has a 
\ce{S}-rich termination. Such asymmetry is reflected in the band structure. 


\subsubsection{\label{AdsBut}Adsorption of butadiene on \ce{MoS2} nanoribbons. }

\begin{figure*}[t]
\includegraphics[scale=0.65,angle=-90]{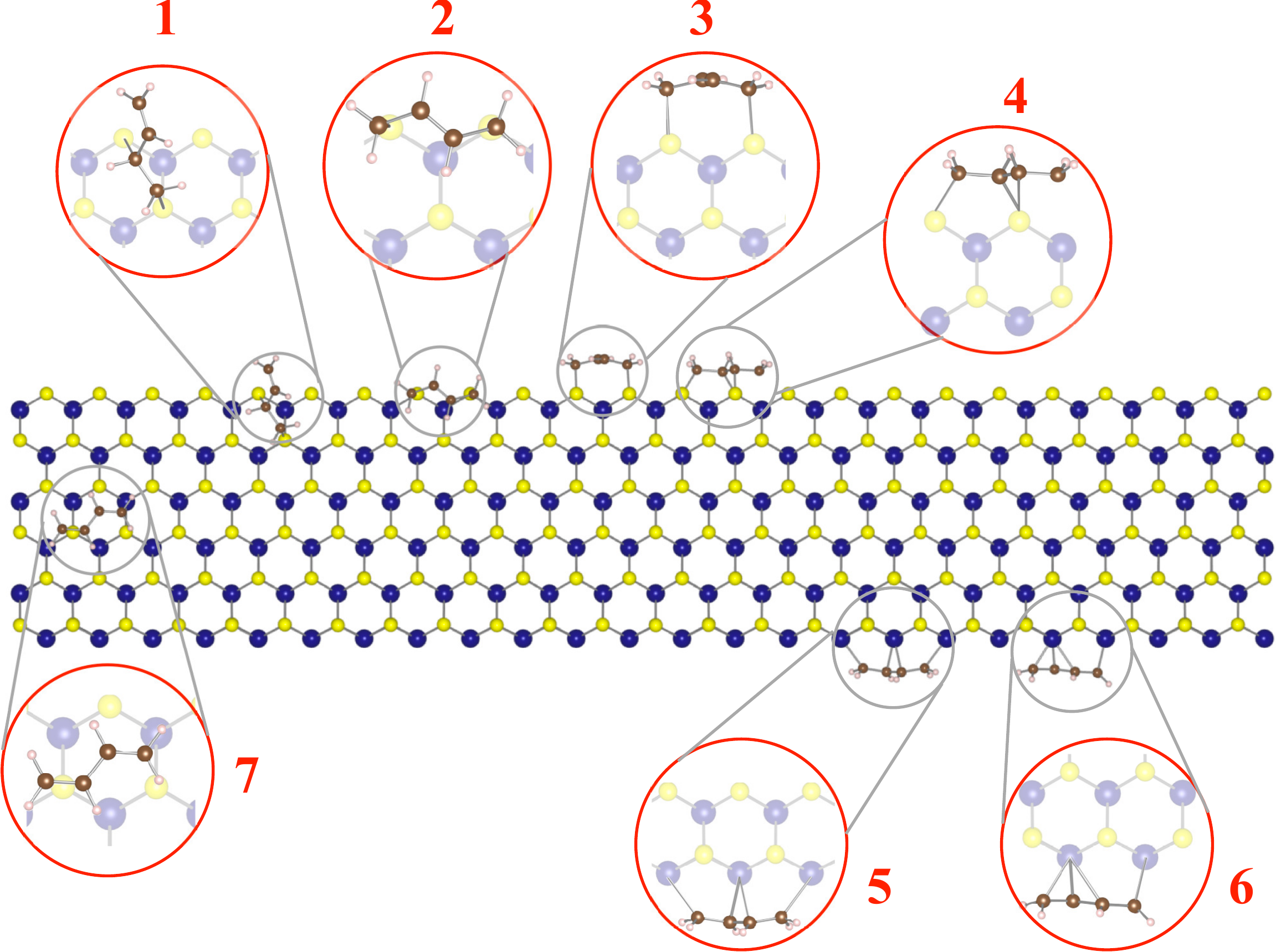}
\centering
\caption{\ce{MoS2}-2H stick-and-ball nanoribbon representation with different butadiene molecule's positions labeled as are represented in table I. This is illustrative representation, since each configuration was calculated in a isolated large supercell. }
\label{fig:stickball}
\end{figure*}

With the assumption that transport will take place along the edges, which is more likely given the results obtained so far, we proceed to find the configurations of butadiene molecules adsorbed on \ce{MoS2} nanoribbons. We start from a selected group of configurations and performed a structural optimization of each configuration. With the optimized geometry we calculated the configuration binding energy (E$_B$), which is given by:

\begin{equation}
E_{B} = E_{Mol.+\ce{MoS2}} - (E_{Mol.}+E_{\ce{MoS2}}),
\end{equation}
where ${{E}_{\text{Mol.}+\text{\ce{MoS2}}}}$ , ${{E}_{\text{\ce{MoS2}}}}$  and ${{E}_{\text{Mol.}}}$  are the total energies of the butdiene molecule adsorbed, in a given configuration, on \ce{MoS2}, \ce{MoS2} nanoribbon and isolated butadiene molecule, respectively. The resulting non-equivalent configurations are showed in Fig.~\ref{fig:stickball}. Probably there are many other adsorption configurations. However, the set of structures presented above is a good sample of possibilities to investigate the transport properties once we considered bulk and edge adsorption sites (on both sides of the ribbon). 

\begin{figure}[h]
\centering
\includegraphics[scale=0.33,angle=0]{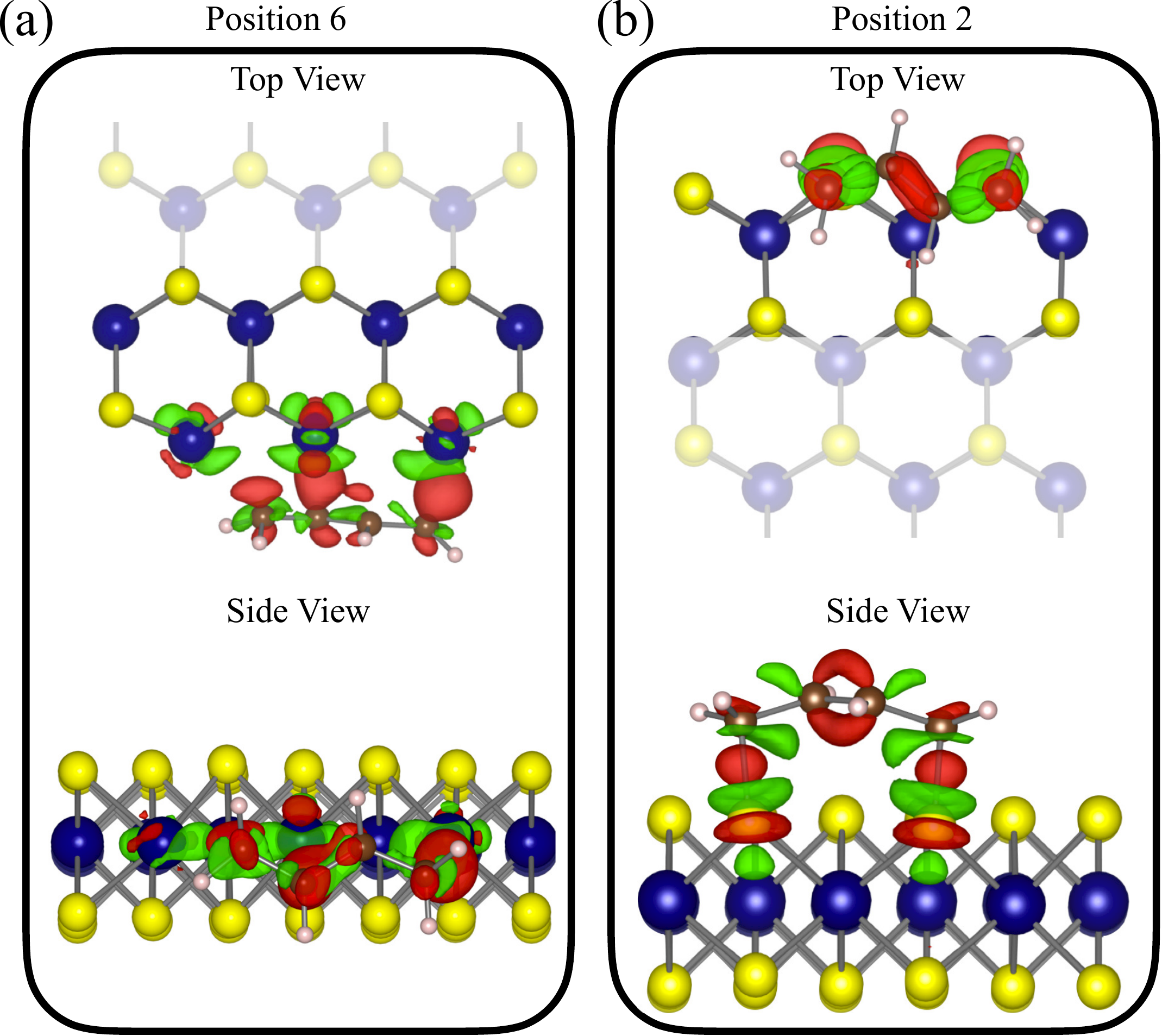}
\caption{\label{charge-transfer} Butadiene-\ce{MoS2} charge transfer. The butadiene molecule is located at the Mo (a) and S (b) termination side. We considered the lowest energy configuration on each termination, according to table~\ref{binding-energy}. The green (red) isosurfaces represents accumulation (depletion) of charge. An isosurface of 0.008 (\si{\elementarycharge/\angstrom^3}) is considered.}
\end{figure}

Table~\ref{binding-energy} presents the binding energies for our sample
adsorption geometries. Our results showed that the two most stable
configurations has the molecule bonded adsorbed on the \ce{Mo}-rich edge.
In addition to these configurations, other four adsorption sites with the
butadiene molecules attached to the \ce{S}-rich border are considered,
including a particular case where the molecule sits just in-between the edges above the ribbon.
These results are in agreement with the charge transfer plot, presented in 
the Fig.~\ref{charge-transfer}, where our zero binding energy reference is
characterized by a broader accumulation region along the edge.

Finally, the molecule adsorbed at the central region of the ribbon (configuration 7) has the highest 
binding energy among all configurations, indicating an energetic preference for edge adsorption. 
However, its absolute value (\SI{0.17}{\electronvolt}) is not small when compared to other systems\cite{PhysRevB.77.125416}, indicating a relatively strong interaction between butadiene molecules and the \ce{MoS2} nanoribbon for any one of the investigated configurations. 

\begin{table}[h]
\small
  \caption{\ Binding energy for the different butadiene positions in \si{\electronvolt}/molecule. The negative sign indicated an exothermic process.}
  \label{binding-energy}
  \centering
  \begin{tabular*}{0.3\textwidth}{@{\extracolsep{\fill}}cc}
    \hline
    Position & Binding energy \\
    \hline
   1 & -0.65  \\ 
   2 & -1.53  \\
   3 & -1.20  \\
   4 & -0.82  \\
   5 & -2.82  \\ 
   6 & -3.70  \\
   7 & -0.17  \\ [1.5ex]
    \hline
  \end{tabular*}
\end{table}

\subsection{Transport properties\label{sec:trans}}
\subsubsection{Pristine nanoribbon.}

\begin{figure}[h]
\includegraphics[scale=0.35,angle=-90]{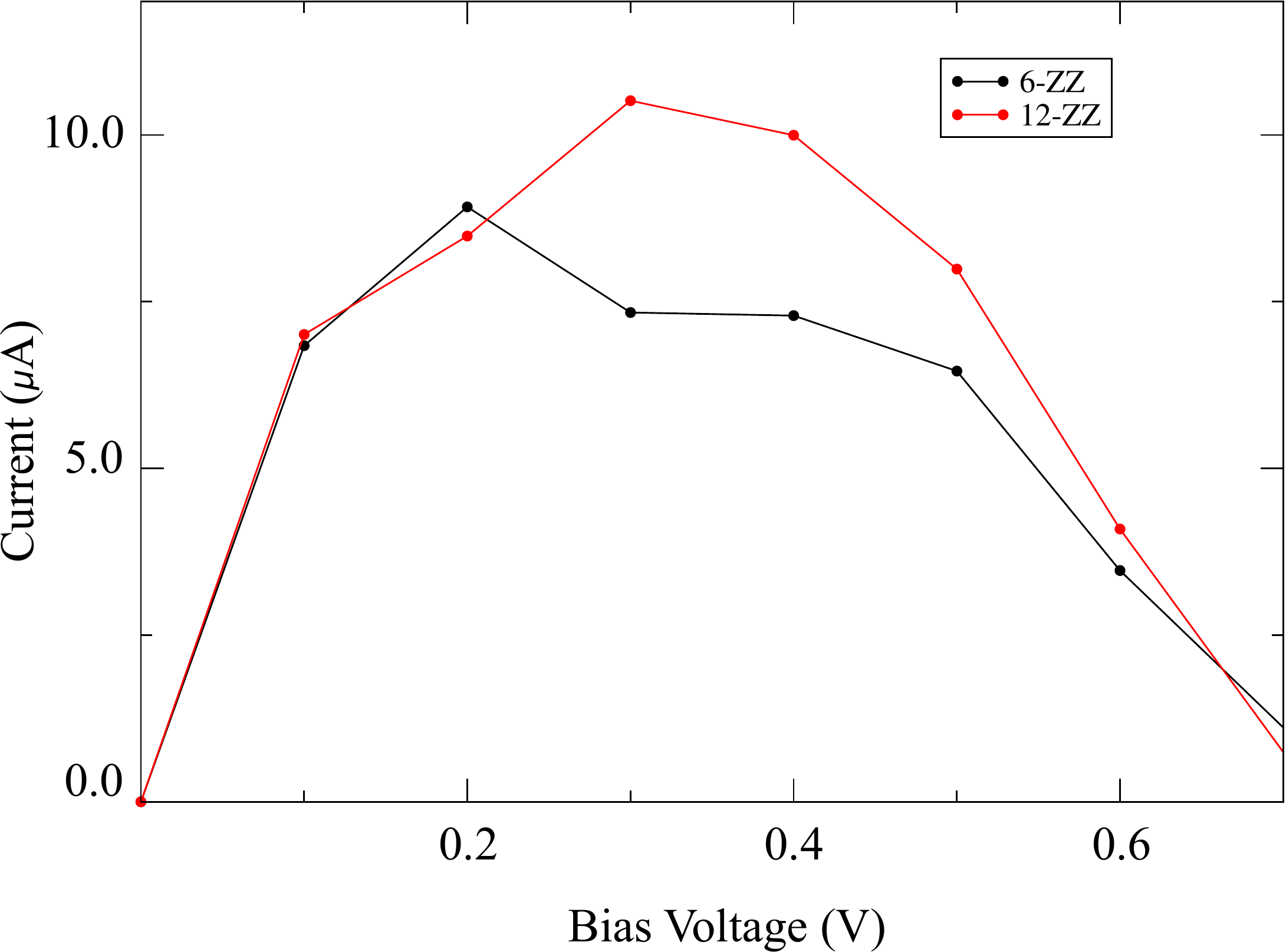}
\centering
\caption{Current as a function of Bias voltage for 6 and 12 zigzag lines. The Negative Differential Resistance is observed in the same bias energy window as reported previously for both cases and the behaviour of the current for both NR's looks similar as well.}
\label{current-pristine}
\end{figure}

In this section, we discuss the transport properties of pristine 
ZZNRs of TMDCs. 
To generate the results presented here we employ a single block of pristine nanoribbon as the scattering region.    
In Fig. \ref{current-pristine} we present the source-drain electric 
current as a function of the bias voltage. Nanoribbons with 6 and 
12 zigzag lines are considered to take into account a width dependence. 
It is worth to note the NDR for 
bias voltages between \num{0.3} and \SI{0.7}{\volt}. Such behavior is introduced by the 
metallic states localized at the edge of the nanoribbons. For bias 
voltages up to \SI{0.3}{\volt}, the edge states of the left and right electrodes 
are almost aligned in energy,  resulting in a high probability for 
the electrons to cross the scattering region. As the bias increases, 
the mismatch between edge channels of both electrodes yields to a decreasing 
in the current, resulting in the NDR behavior. This feature is in agreement 
with the previous results\cite{0022-3727-49-24-245304}.

\subsection{Lowest energy configuration}

\begin{figure}[h]
\hspace*{-1.4cm}
\includegraphics[scale=0.45,angle=0]{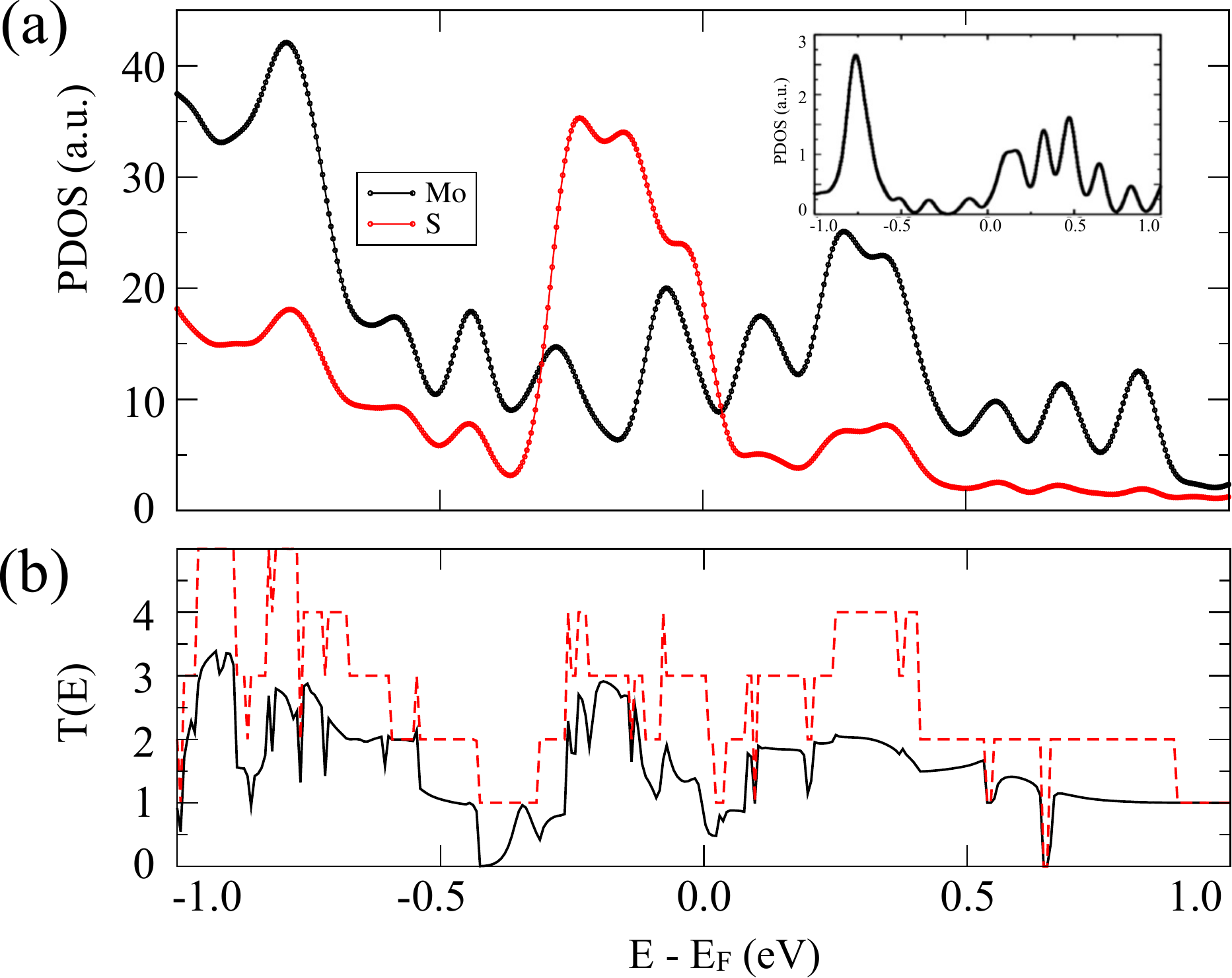}
\centering
\caption{\label{pdos_T}Projected density of states (PDOS) (a), and transmission coefficients (T) (b), for the most stable configuration (position 6). In (a) the inset shows the PDOS for the molecule in the same energy window used for both PDOS and T(E) while in (b) the dashed red curve is the transmission T(E) for the pristine structure. The Fermi level is set to zero. }
\end{figure}

Among the seven adsorption sites aforementioned in section 
\ref{AdsBut}, we demonstrate here the effects in the transport properties 
of TMDCs ZZNRs generated by a single 
butadiene molecule adsorbed in its most stable configuration. 
To generate these results we consider a scattering region 
comprised of a single ZZNR block with 7 unit cells and one 
butadiene molecule adsorbed in the position 6.  
In this structure, the molecule is located on the \ce{Mo}-rich 
edge, and its high binding energy (\SI{3.7}{\electronvolt}) demonstrates 
a strong interaction. Fig.~\ref{pdos_T}, upper panel, 
shows the local density of states (LDOS), where we've previously shown that those states live along the nanoribbon borders, meaning that states coming from the molecule (inset), interact with the $\ce{Mo}$ orbitals localized at the edge. As a result, we can see how the transmission curve (T(E)) have changed with respect to that one of the pristine case, red and black respectively.
Fig.~\ref{pdos_T}, bottom panel, shows T(E). The solid (dashed) line 
corresponds to configuration 6 (pristine nanoribbon). The butadiene 
molecule generates a decreasing of T(E) around the Fermi level by almost 
one transport channel.

\subsubsection{Large scale disordered system.}

Up to this point, the obtained results suggest the next stage:  since 
ZZNRs present (metallic) edge states in an energy window around the 
Fermi level, what is the influence in the transport properties if
butadiene molecules used in the exfoliation process anchor at the 
edges of large-scale ZZNRs? At first glance, these
interactions will change the electronic structure, and consequently the
electronic transport, as indicated in the previous section. The chemical rearrangement experienced by the sample
could be related to the existence of quasi-localized states, acting as
source of electron backscattering processes and leading to a drop 
in the transmission. Another possibility is the presence of stronger
localized states which could
lead to a different electronic transport trend.

\begin{figure}[h]
\includegraphics[angle=-90,scale=0.7]{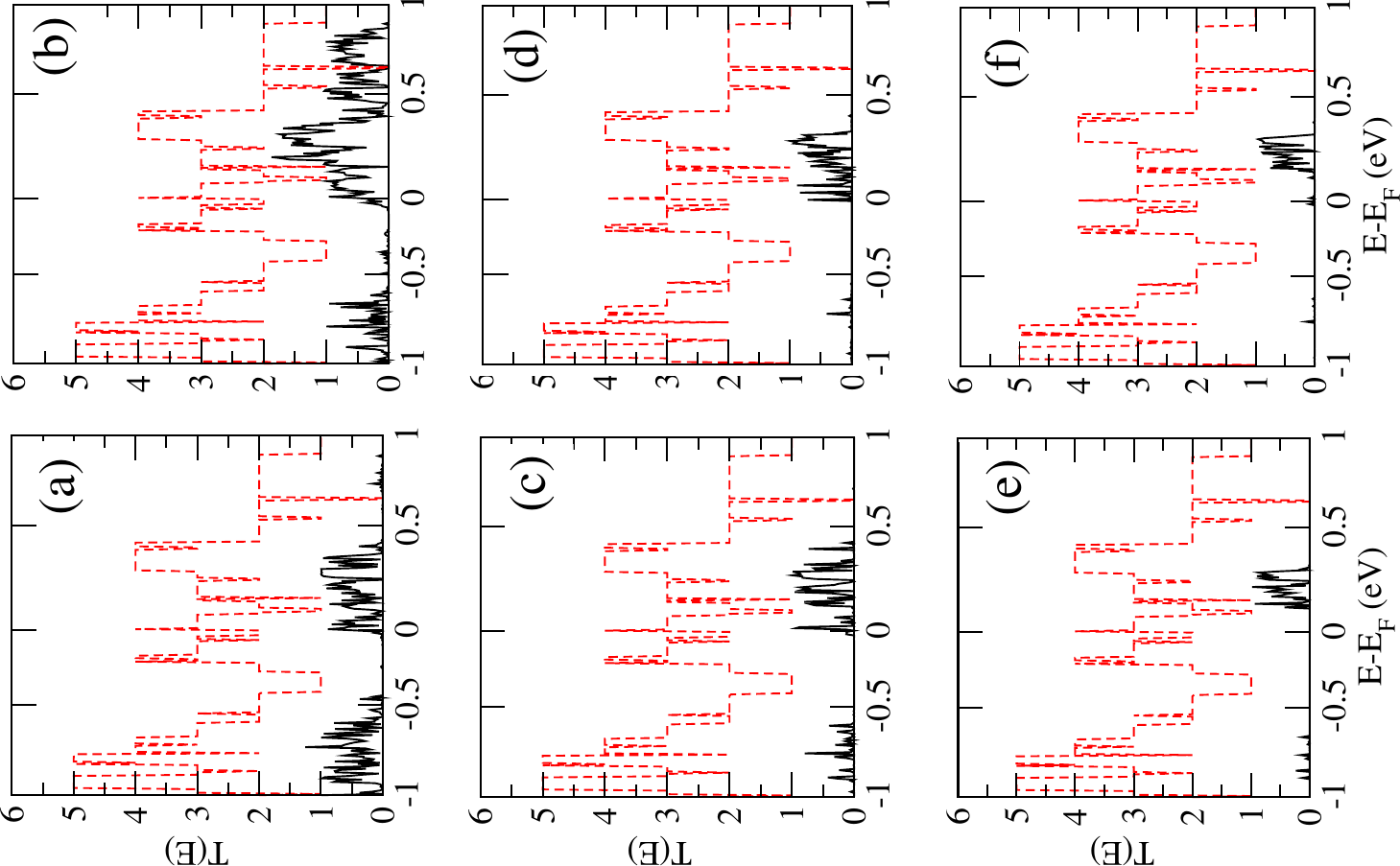}
\centering
\caption{\label{T-molecule}Transport coefficients for (a) 5, (b) 10, (c) 20, (d) 30, (e) 40 and (f) 50 molecules attached to the NNR (black curves). For sake of comparison we also show the transmission calculated for the pristine structure (red curve). Each of them was calculated under a 0.1 V bias.}
\end{figure}

The previous section demonstrates that a single adsorbed molecule already 
significativelly modify the transport properties of TMDC nanoribbons. In 
this section, we are interested in to understand the effects of several 
butadiene molecules in large scale nanoribbons. 
Our results are obtained by using the previous arrangements as building blocks to increase the size of our scattering region along the transport direction, as we've mentioned, this is possible with the application of a decimation process\cite{recursive1,recursive2}. With this procedure, we obtain the transmission function shown in Fig.~\ref{T-molecule}, where we have different curves for different quantities of molecules attached to the ribbon and are presented as an average over thirty different distributions with the same number of molecules, the dashed line represents the pristine case and the others show how the values of the transmission are getting smaller as the number of molecules becomes larger. The scatter length can be estimated by the following relation

\begin{equation}
L_{Total}(n)=n*L_{Molecule}+(50-n)*L_{Pristine}
\end{equation}

\begin{figure}[h]
\includegraphics[angle=0,scale=0.35]{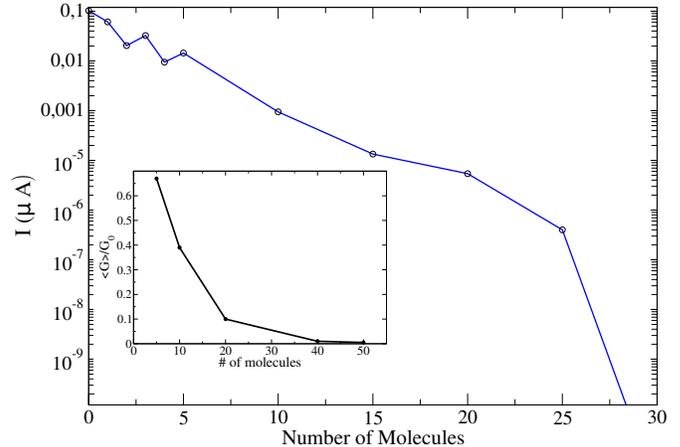}
\centering
\caption{\label{all-molecule} Current in $\mu A$ for different number of molecules attached to the NR for an applied bias of \SI{0.1}{\volt}, the values are present in a log scale in order to show how the current decreases by orders of magnitude once the number of molecules increases, besides we show how the average conductance $<G>/G_0$ ($G_0$ being the conductance quantum) at the Fermi level decays.}
\end{figure}

Where $L_{Total}$ is the total length of the scattering region, $n$ the number of molecules, $L_{Molecule}$ is the length of a piece of NR with one molecule attached (which is the same for all the molecules) and $L_{Pristine}$ is the length of molecule-free pieces of NR which are used as buffer regions. Having these values, for 5 molecules the total length is of \SI{574.2}{\angstrom}, a number which increases ranging values up to \SI{1052.7}{\angstrom} for 30 molecules. 


In Fig.~\ref{all-molecule}, we show the current obtained with an applied bias voltage of \SI{0.1}{\volt}; again this figure shows a current decreasing with the increase on the number of molecules. In the inset we also show the conductance calculated as an average, given the $T(E)$ we know that if $\mu$ represents the chemical potential we can write
\begin{equation}
G(\mu)=\int dE \thinspace T(E) {df_{FD}(E'-\mu)\over dE'}\bigg|_{E'=E},
\end{equation}
where the conductance is calculated in units of \si{2\elementarycharge^2/h}. The observed values obtained for the current are supported by the average conductance curve, which has a decaying behaviour suggesting that butadiene molecules yield to a localization regime once we are working in the nanometer scale.\\

On the experimental side, processing residues and imperfect electrical contacts play an important role when it comes to determine regimes at which the transport measurements are made \cite{doi:10.1146/annurev-physchem-050317-021353,Bhandari_2017}. The use of \ce{Si}/\ce{SiO2} as substrate lead to a localization up to room temperature, which is related to the disorder arising from randomly distributed charges at the interface between the sample and substrate, in others words, this suggests that the electronic behaviour could be enhanced by a careful engineering of the substrate made for the sample \cite{doi:10.1021/nn202852j}.
  
The transport calculations show a correlation between the existence of the quasi-localized states and the decreasing in the transport coefficients, as usual, this has to do with backscattering processes which lead to a suppressed transmission, where is evident how the transmission decreases as there are more molecules attached to the nanoribbon then acting as source of scattering processes~\cite{PhysRevLett.102.096803,PhysRevB.81.085427,doi:10.1021/nl802798q}. Previous experimental works pointed out the charged impurities and localized states are very important in exfoliated samples, different sources give rise to scattering processes and then it's hard to find an accurate transport mechanism for this cases. The existence of localized states usually makes difficult a ballistic transport but it's been observed that this is possible when the sample is highly doped where band transport can be achieved \cite{doi:10.1021/nl401916s,doi:10.1021/nl4046922,doi:10.1063/1.4803920}.\\

\section{Conclusions}

In this work, we investigated the influence of butadiene molecules in the transport properties of \ce{MoS2} zig-zag nanoribbons with metallic edge states. Our simulations took into account the effects of disorder and different coverages in large-scale systems (\si{\approx \micro\metre}). Our principal motivation is the experimental use of butadiene molecules in the dimensionality reduction process of \ce{MoS2}, recently published in the literature\cite{doi:10.1002/chem.201502303}. Our results allow to infer that \ce{MoS2} nanoribbons containing adsorbed butadiene molecules lost its metallic behavior due to backscattering process. Such a behavior highly contrasts with pristine TMDCs nanoribbons. Furthermore, the relatively high absolute value for the binding energy indicates that a cleaning process (removing butadiene molecules from \ce{MoS2}) is a hard task. Thus, the mechanisms used to reduce the dimensionality of \ce{MoS2} considerably modify its transport properties.


\section{Conflict of Interest}
The authors declare that there are no conflicts of interest.
\section{Acknowledgements}

This work was supported by the Brazilian agencies \textit{Funda\c{c}\~ao de Amparo \`a Pesquisa do Estado de S\~ao Paulo (FAPESP)} (grants 16/14011-2 and 17/02317-2), CAPES and INCT/CNPQ. We would like to acknowledge computing time to the UFABC, and SAMPA group for the computational facilities. The authors also acknowledge the National Laboratory for Scientific Computing (LNCC/MCTI, Brazil) for providing HPC resources of the SDumont supercomputer, which have contributed to the research results reported within this paper. URL: http://sdumont.lncc.br



\balance


\bibliography{bibliography} 

\providecommand*{\mcitethebibliography}{\thebibliography}
\csname @ifundefined\endcsname{endmcitethebibliography}
{\let\endmcitethebibliography\endthebibliography}{}
\begin{mcitethebibliography}{58}
\providecommand*{\natexlab}[1]{#1}
\providecommand*{\mciteSetBstSublistMode}[1]{}
\providecommand*{\mciteSetBstMaxWidthForm}[2]{}
\providecommand*{\mciteBstWouldAddEndPuncttrue}
  {\def\EndOfBibitem{\unskip.}}
\providecommand*{\mciteBstWouldAddEndPunctfalse}
  {\let\EndOfBibitem\relax}
\providecommand*{\mciteSetBstMidEndSepPunct}[3]{}
\providecommand*{\mciteSetBstSublistLabelBeginEnd}[3]{}
\providecommand*{\EndOfBibitem}{}
\mciteSetBstSublistMode{f}
\mciteSetBstMaxWidthForm{subitem}
{(\emph{\alph{mcitesubitemcount}})}
\mciteSetBstSublistLabelBeginEnd{\mcitemaxwidthsubitemform\space}
{\relax}{\relax}

\bibitem[Novoselov \emph{et~al.}(2004)Novoselov, Geim, Morozov, Jiang, Zhang,
  Dubonos, Grigorieva, and Firsov]{Novoselov666}
K.~S. Novoselov, A.~K. Geim, S.~V. Morozov, D.~Jiang, Y.~Zhang, S.~V. Dubonos,
  I.~V. Grigorieva and A.~A. Firsov, \emph{Science}, 2004, \textbf{306},
  666--669\relax
\mciteBstWouldAddEndPuncttrue
\mciteSetBstMidEndSepPunct{\mcitedefaultmidpunct}
{\mcitedefaultendpunct}{\mcitedefaultseppunct}\relax
\EndOfBibitem
\bibitem[Novoselov \emph{et~al.}(2016)Novoselov, Mishchenko, Carvalho, and
  Castro~Neto]{2Dera}
K.~S. Novoselov, A.~Mishchenko, A.~Carvalho and A.~H. Castro~Neto,
  \emph{Science}, 2016, \textbf{353}, aac9439\relax
\mciteBstWouldAddEndPuncttrue
\mciteSetBstMidEndSepPunct{\mcitedefaultmidpunct}
{\mcitedefaultendpunct}{\mcitedefaultseppunct}\relax
\EndOfBibitem
\bibitem[Novoselov \emph{et~al.}(2005)Novoselov, Geim, Morozov, Jiang,
  Katsnelson, Grigorieva, Dubonos, and Firsov]{Novoselov2005}
K.~S. Novoselov, A.~K. Geim, S.~V. Morozov, D.~Jiang, M.~I. Katsnelson, I.~V.
  Grigorieva, S.~V. Dubonos and A.~A. Firsov, \emph{Nature}, 2005,
  \textbf{438}, 197--200\relax
\mciteBstWouldAddEndPuncttrue
\mciteSetBstMidEndSepPunct{\mcitedefaultmidpunct}
{\mcitedefaultendpunct}{\mcitedefaultseppunct}\relax
\EndOfBibitem
\bibitem[Kim \emph{et~al.}(2009)Kim, Zhao, Jang, Lee, Kim, Kim, Ahn, Kim, Choi,
  and Hong]{Kim2009}
K.~S. Kim, Y.~Zhao, H.~Jang, S.~Y. Lee, J.~M. Kim, K.~S. Kim, J.-H. Ahn,
  P.~Kim, J.-Y. Choi and B.~H. Hong, \emph{Nature}, 2009, \textbf{457}, 706 EP
  --\relax
\mciteBstWouldAddEndPuncttrue
\mciteSetBstMidEndSepPunct{\mcitedefaultmidpunct}
{\mcitedefaultendpunct}{\mcitedefaultseppunct}\relax
\EndOfBibitem
\bibitem[Balandin \emph{et~al.}(2008)Balandin, Ghosh, Bao, Calizo,
  Teweldebrhan, Miao, and Lau]{doi:10.1021/nl0731872}
A.~A. Balandin, S.~Ghosh, W.~Bao, I.~Calizo, D.~Teweldebrhan, F.~Miao and C.~N.
  Lau, \emph{Nano Letters}, 2008, \textbf{8}, 902--907\relax
\mciteBstWouldAddEndPuncttrue
\mciteSetBstMidEndSepPunct{\mcitedefaultmidpunct}
{\mcitedefaultendpunct}{\mcitedefaultseppunct}\relax
\EndOfBibitem
\bibitem[Falkovsky(2008)]{Falkovsky_2008}
L.~A. Falkovsky, \emph{Journal of Physics: Conference Series}, 2008,
  \textbf{129}, 012004\relax
\mciteBstWouldAddEndPuncttrue
\mciteSetBstMidEndSepPunct{\mcitedefaultmidpunct}
{\mcitedefaultendpunct}{\mcitedefaultseppunct}\relax
\EndOfBibitem
\bibitem[Castro~Neto \emph{et~al.}(2009)Castro~Neto, Guinea, Peres, Novoselov,
  and Geim]{RevModPhys.81.109}
A.~H. Castro~Neto, F.~Guinea, N.~M.~R. Peres, K.~S. Novoselov and A.~K. Geim,
  \emph{Rev. Mod. Phys.}, 2009, \textbf{81}, 109--162\relax
\mciteBstWouldAddEndPuncttrue
\mciteSetBstMidEndSepPunct{\mcitedefaultmidpunct}
{\mcitedefaultendpunct}{\mcitedefaultseppunct}\relax
\EndOfBibitem
\bibitem[Han \emph{et~al.}(2015)Han, Bhatia, and Kim]{Han2015}
S.~A. Han, R.~Bhatia and S.-W. Kim, \emph{Nano Convergence}, 2015, \textbf{2},
  17\relax
\mciteBstWouldAddEndPuncttrue
\mciteSetBstMidEndSepPunct{\mcitedefaultmidpunct}
{\mcitedefaultendpunct}{\mcitedefaultseppunct}\relax
\EndOfBibitem
\bibitem[Korm\'{a}nyos \emph{et~al.}(2015)Korm\'{a}nyos, Burkard, Gmitra,
  Fabian, Z\'{o}lyomi, Drummond, and Fal'ko]{2053-1583-2-2-022001}
A.~Korm\'{a}nyos, G.~Burkard, M.~Gmitra, J.~Fabian, V.~Z\'{o}lyomi, N.~D.
  Drummond and V.~Fal'ko, \emph{2D Materials}, 2015, \textbf{2}, 022001\relax
\mciteBstWouldAddEndPuncttrue
\mciteSetBstMidEndSepPunct{\mcitedefaultmidpunct}
{\mcitedefaultendpunct}{\mcitedefaultseppunct}\relax
\EndOfBibitem
\bibitem[Ataca and Ciraci(2011)]{Ataca-1}
C.~Ataca and S.~Ciraci, \emph{The Journal of Physical Chemistry C}, 2011,
  \textbf{115}, 13303--13311\relax
\mciteBstWouldAddEndPuncttrue
\mciteSetBstMidEndSepPunct{\mcitedefaultmidpunct}
{\mcitedefaultendpunct}{\mcitedefaultseppunct}\relax
\EndOfBibitem
\bibitem[Qian \emph{et~al.}(2014)Qian, Liu, Fu, and Li]{Qian1344}
X.~Qian, J.~Liu, L.~Fu and J.~Li, \emph{Science}, 2014, \textbf{346},
  1344--1347\relax
\mciteBstWouldAddEndPuncttrue
\mciteSetBstMidEndSepPunct{\mcitedefaultmidpunct}
{\mcitedefaultendpunct}{\mcitedefaultseppunct}\relax
\EndOfBibitem
\bibitem[Chhowalla \emph{et~al.}(2013)Chhowalla, Shin, Eda, Li, Loh, and
  Zhang]{chem2}
M.~Chhowalla, H.~S. Shin, G.~Eda, L.-J. Li, K.~P. Loh and H.~Zhang,
  \emph{Nature Chemistry}, 2013, \textbf{5}, 263 EP --\relax
\mciteBstWouldAddEndPuncttrue
\mciteSetBstMidEndSepPunct{\mcitedefaultmidpunct}
{\mcitedefaultendpunct}{\mcitedefaultseppunct}\relax
\EndOfBibitem
\bibitem[An \emph{et~al.}(2018)An, Jiao, Hou, Wang, Wu, Liu, Chen, Wang, and
  Wang]{An_2018}
Y.~An, J.~Jiao, Y.~Hou, H.~Wang, R.~Wu, C.~Liu, X.~Chen, T.~Wang and K.~Wang,
  \emph{Journal of Physics: Condensed Matter}, 2018, \textbf{31}, 065301\relax
\mciteBstWouldAddEndPuncttrue
\mciteSetBstMidEndSepPunct{\mcitedefaultmidpunct}
{\mcitedefaultendpunct}{\mcitedefaultseppunct}\relax
\EndOfBibitem
\bibitem[{An} \emph{et~al.}(2018){An}, {Sun}, {Zhang}, {Jiao}, {Wu}, {Wang},
  and {Wang}]{8443152}
Y.~{An}, Y.~{Sun}, M.~{Zhang}, J.~{Jiao}, D.~{Wu}, T.~{Wang} and K.~{Wang},
  \emph{IEEE Transactions on Electron Devices}, 2018, \textbf{65},
  4646--4651\relax
\mciteBstWouldAddEndPuncttrue
\mciteSetBstMidEndSepPunct{\mcitedefaultmidpunct}
{\mcitedefaultendpunct}{\mcitedefaultseppunct}\relax
\EndOfBibitem
\bibitem[An \emph{et~al.}(2017)An, Sun, Jiao, Zhang, Wang, Chen, Wu, Wang, Fu,
  and Jiao]{AN201743}
Y.~An, Y.~Sun, J.~Jiao, M.~Zhang, K.~Wang, X.~Chen, D.~Wu, T.~Wang, Z.~Fu and
  Z.~Jiao, \emph{Organic Electronics}, 2017, \textbf{50}, 43 -- 47\relax
\mciteBstWouldAddEndPuncttrue
\mciteSetBstMidEndSepPunct{\mcitedefaultmidpunct}
{\mcitedefaultendpunct}{\mcitedefaultseppunct}\relax
\EndOfBibitem
\bibitem[Zhang \emph{et~al.}(2017)Zhang, An, Sun, Wu, Chen, Wang, Xu, and
  Wang]{C7CP02201H}
M.~Zhang, Y.~An, Y.~Sun, D.~Wu, X.~Chen, T.~Wang, G.~Xu and K.~Wang,
  \emph{Phys. Chem. Chem. Phys.}, 2017, \textbf{19}, 17210--17215\relax
\mciteBstWouldAddEndPuncttrue
\mciteSetBstMidEndSepPunct{\mcitedefaultmidpunct}
{\mcitedefaultendpunct}{\mcitedefaultseppunct}\relax
\EndOfBibitem
\bibitem[An \emph{et~al.}(2016)An, Zhang, Wu, Fu, and Wang]{C6TC04327E}
Y.~An, M.~Zhang, D.~Wu, Z.~Fu and K.~Wang, \emph{J. Mater. Chem. C}, 2016,
  \textbf{4}, 10962--10966\relax
\mciteBstWouldAddEndPuncttrue
\mciteSetBstMidEndSepPunct{\mcitedefaultmidpunct}
{\mcitedefaultendpunct}{\mcitedefaultseppunct}\relax
\EndOfBibitem
\bibitem[An \emph{et~al.}(2018)An, Jiao, Hou, Wang, Wu, Wang, Fu, Xu, and
  Wu]{C8CP04272A}
Y.~An, J.~Jiao, Y.~Hou, H.~Wang, D.~Wu, T.~Wang, Z.~Fu, G.~Xu and R.~Wu,
  \emph{Phys. Chem. Chem. Phys.}, 2018, \textbf{20}, 21552--21556\relax
\mciteBstWouldAddEndPuncttrue
\mciteSetBstMidEndSepPunct{\mcitedefaultmidpunct}
{\mcitedefaultendpunct}{\mcitedefaultseppunct}\relax
\EndOfBibitem
\bibitem[Chia \emph{et~al.}(2015)Chia, Ambrosi, Sofer, Luxa, and
  Pumera]{doi:10.1021/acsnano.5b00501}
X.~Chia, A.~Ambrosi, Z.~Sofer, J.~Luxa and M.~Pumera, \emph{ACS Nano}, 2015,
  \textbf{9}, 5164--5179\relax
\mciteBstWouldAddEndPuncttrue
\mciteSetBstMidEndSepPunct{\mcitedefaultmidpunct}
{\mcitedefaultendpunct}{\mcitedefaultseppunct}\relax
\EndOfBibitem
\bibitem[Zhang and Zhang(2017)]{C7TC01088E}
G.~Zhang and Y.-W. Zhang, \emph{J. Mater. Chem. C}, 2017, \textbf{5},
  7684--7698\relax
\mciteBstWouldAddEndPuncttrue
\mciteSetBstMidEndSepPunct{\mcitedefaultmidpunct}
{\mcitedefaultendpunct}{\mcitedefaultseppunct}\relax
\EndOfBibitem
\bibitem[Akama \emph{et~al.}(2017)Akama, Okita, Nagai, Li, Kaneko, and
  Kato]{Akama2017}
T.~Akama, W.~Okita, R.~Nagai, C.~Li, T.~Kaneko and T.~Kato, \emph{Scientific
  Reports}, 2017, \textbf{7}, 11967\relax
\mciteBstWouldAddEndPuncttrue
\mciteSetBstMidEndSepPunct{\mcitedefaultmidpunct}
{\mcitedefaultendpunct}{\mcitedefaultseppunct}\relax
\EndOfBibitem
\bibitem[Pan and Zhang(2012)]{C2JM15906F}
H.~Pan and Y.-W. Zhang, \emph{J. Mater. Chem.}, 2012, \textbf{22},
  7280--7290\relax
\mciteBstWouldAddEndPuncttrue
\mciteSetBstMidEndSepPunct{\mcitedefaultmidpunct}
{\mcitedefaultendpunct}{\mcitedefaultseppunct}\relax
\EndOfBibitem
\bibitem[Ataca \emph{et~al.}(2012)Ataca, Sahin, and Ciraci]{Ataca-2}
C.~Ataca, H.~Sahin and S.~Ciraci, \emph{The Journal of Physical Chemistry C},
  2012, \textbf{116}, 8983--8999\relax
\mciteBstWouldAddEndPuncttrue
\mciteSetBstMidEndSepPunct{\mcitedefaultmidpunct}
{\mcitedefaultendpunct}{\mcitedefaultseppunct}\relax
\EndOfBibitem
\bibitem[Liu \emph{et~al.}(2012)Liu, Neal, and Ye]{doi:10.1021/nn303513c}
H.~Liu, A.~T. Neal and P.~D. Ye, \emph{ACS Nano}, 2012, \textbf{6},
  8563--8569\relax
\mciteBstWouldAddEndPuncttrue
\mciteSetBstMidEndSepPunct{\mcitedefaultmidpunct}
{\mcitedefaultendpunct}{\mcitedefaultseppunct}\relax
\EndOfBibitem
\bibitem[Yoon \emph{et~al.}(2011)Yoon, Ganapathi, and
  Salahuddin]{doi:10.1021/nl2018178}
Y.~Yoon, K.~Ganapathi and S.~Salahuddin, \emph{Nano Letters}, 2011,
  \textbf{11}, 3768--3773\relax
\mciteBstWouldAddEndPuncttrue
\mciteSetBstMidEndSepPunct{\mcitedefaultmidpunct}
{\mcitedefaultendpunct}{\mcitedefaultseppunct}\relax
\EndOfBibitem
\bibitem[An \emph{et~al.}(2016)An, Zhang, Da, Fu, Jiao, and
  Liu]{0022-3727-49-24-245304}
Y.~An, M.~Zhang, H.~Da, Z.~Fu, Z.~Jiao and Z.~Liu, \emph{Journal of Physics D:
  Applied Physics}, 2016, \textbf{49}, 245304\relax
\mciteBstWouldAddEndPuncttrue
\mciteSetBstMidEndSepPunct{\mcitedefaultmidpunct}
{\mcitedefaultendpunct}{\mcitedefaultseppunct}\relax
\EndOfBibitem
\bibitem[Coleman \emph{et~al.}(2011)Coleman, Lotya, O{\textquoteright}Neill,
  Bergin, King, Khan, Young, Gaucher, De, Smith, Shvets, Arora, Stanton, Kim,
  Lee, Kim, Duesberg, Hallam, Boland, Wang, Donegan, Grunlan, Moriarty,
  Shmeliov, Nicholls, Perkins, Grieveson, Theuwissen, McComb, Nellist, and
  Nicolosi]{chem1}
J.~N. Coleman, M.~Lotya, A.~O{\textquoteright}Neill, S.~D. Bergin, P.~J. King,
  U.~Khan, K.~Young, A.~Gaucher, S.~De, R.~J. Smith, I.~V. Shvets, S.~K. Arora,
  G.~Stanton, H.-Y. Kim, K.~Lee, G.~T. Kim, G.~S. Duesberg, T.~Hallam, J.~J.
  Boland, J.~J. Wang, J.~F. Donegan, J.~C. Grunlan, G.~Moriarty, A.~Shmeliov,
  R.~J. Nicholls, J.~M. Perkins, E.~M. Grieveson, K.~Theuwissen, D.~W. McComb,
  P.~D. Nellist and V.~Nicolosi, \emph{Science}, 2011, \textbf{331},
  568--571\relax
\mciteBstWouldAddEndPuncttrue
\mciteSetBstMidEndSepPunct{\mcitedefaultmidpunct}
{\mcitedefaultendpunct}{\mcitedefaultseppunct}\relax
\EndOfBibitem
\bibitem[Gon{\c{c}}alves \emph{et~al.}(2015)Gon{\c{c}}alves, Fiel, Soares,
  Schreiner, Silva, and Leite]{doi:10.1002/chem.201502303}
R.~H. Gon{\c{c}}alves, R.~Fiel, M.~R.~S. Soares, W.~H. Schreiner, C.~M.~P.
  Silva and E.~R. Leite, \emph{Chemistry - A European Journal}, 2015,
  \textbf{21}, 15583--15588\relax
\mciteBstWouldAddEndPuncttrue
\mciteSetBstMidEndSepPunct{\mcitedefaultmidpunct}
{\mcitedefaultendpunct}{\mcitedefaultseppunct}\relax
\EndOfBibitem
\bibitem[Yue \emph{et~al.}(2013)Yue, Shao, Chang, and Li]{Yue2013}
Q.~Yue, Z.~Shao, S.~Chang and J.~Li, \emph{Nanoscale Research Letters}, 2013,
  \textbf{8}, 425\relax
\mciteBstWouldAddEndPuncttrue
\mciteSetBstMidEndSepPunct{\mcitedefaultmidpunct}
{\mcitedefaultendpunct}{\mcitedefaultseppunct}\relax
\EndOfBibitem
\bibitem[Yang \emph{et~al.}(2012)Yang, Li, Zhang, Tao, and
  Chen]{doi:10.1021/jp2097026}
S.~Yang, D.~Li, T.~Zhang, Z.~Tao and J.~Chen, \emph{The Journal of Physical
  Chemistry C}, 2012, \textbf{116}, 1307--1312\relax
\mciteBstWouldAddEndPuncttrue
\mciteSetBstMidEndSepPunct{\mcitedefaultmidpunct}
{\mcitedefaultendpunct}{\mcitedefaultseppunct}\relax
\EndOfBibitem
\bibitem[Ghuman \emph{et~al.}(2015)Ghuman, Yadav, and
  Singh]{doi:10.1021/jp510899m}
K.~K. Ghuman, S.~Yadav and C.~V. Singh, \emph{The Journal of Physical Chemistry
  C}, 2015, \textbf{119}, 6518--6529\relax
\mciteBstWouldAddEndPuncttrue
\mciteSetBstMidEndSepPunct{\mcitedefaultmidpunct}
{\mcitedefaultendpunct}{\mcitedefaultseppunct}\relax
\EndOfBibitem
\bibitem[Cai \emph{et~al.}(2015)Cai, Ke, Zhang, and
  Zhang]{doi:10.1021/jp510863p}
Y.~Cai, Q.~Ke, G.~Zhang and Y.-W. Zhang, \emph{The Journal of Physical
  Chemistry C}, 2015, \textbf{119}, 3102--3110\relax
\mciteBstWouldAddEndPuncttrue
\mciteSetBstMidEndSepPunct{\mcitedefaultmidpunct}
{\mcitedefaultendpunct}{\mcitedefaultseppunct}\relax
\EndOfBibitem
\bibitem[Rodríguez and Albanesi(2019)]{C8CP05093G}
S.~J. Rodríguez and E.~A. Albanesi, \emph{Phys. Chem. Chem. Phys.}, 2019,
  \textbf{21}, 597--606\relax
\mciteBstWouldAddEndPuncttrue
\mciteSetBstMidEndSepPunct{\mcitedefaultmidpunct}
{\mcitedefaultendpunct}{\mcitedefaultseppunct}\relax
\EndOfBibitem
\bibitem[Shukla \emph{et~al.}(2017)Shukla, Wärnå, Jena, Grigoriev, and
  Ahuja]{doi:10.1021/acs.jpcc.7b09552}
V.~Shukla, J.~Wärnå, N.~K. Jena, A.~Grigoriev and R.~Ahuja, \emph{The Journal
  of Physical Chemistry C}, 2017, \textbf{121}, 26869--26876\relax
\mciteBstWouldAddEndPuncttrue
\mciteSetBstMidEndSepPunct{\mcitedefaultmidpunct}
{\mcitedefaultendpunct}{\mcitedefaultseppunct}\relax
\EndOfBibitem
\bibitem[Huang and Cho(2009)]{doi:10.1021/jp807705y}
M.~Huang and K.~Cho, \emph{The Journal of Physical Chemistry C}, 2009,
  \textbf{113}, 5238--5243\relax
\mciteBstWouldAddEndPuncttrue
\mciteSetBstMidEndSepPunct{\mcitedefaultmidpunct}
{\mcitedefaultendpunct}{\mcitedefaultseppunct}\relax
\EndOfBibitem
\bibitem[Ataca \emph{et~al.}(2011)Ataca, Şahin, Aktürk, and
  Ciraci]{doi:10.1021/jp1115146}
C.~Ataca, H.~Şahin, E.~Aktürk and S.~Ciraci, \emph{The Journal of Physical
  Chemistry C}, 2011, \textbf{115}, 3934--3941\relax
\mciteBstWouldAddEndPuncttrue
\mciteSetBstMidEndSepPunct{\mcitedefaultmidpunct}
{\mcitedefaultendpunct}{\mcitedefaultseppunct}\relax
\EndOfBibitem
\bibitem[Hohenberg and Kohn(1964)]{DFT1}
P.~Hohenberg and W.~Kohn, \emph{Physical Review}, 1964, \textbf{136},
  B864--B871\relax
\mciteBstWouldAddEndPuncttrue
\mciteSetBstMidEndSepPunct{\mcitedefaultmidpunct}
{\mcitedefaultendpunct}{\mcitedefaultseppunct}\relax
\EndOfBibitem
\bibitem[Schleder \emph{et~al.}(2019.
  https://doi.org/10.1088/2515-7639/ab084b)Schleder, Padilha, Acosta, Costa,
  and Fazzio]{review}
G.~R. Schleder, A.~C.~M. Padilha, C.~M. Acosta, M.~Costa and A.~Fazzio,
  \emph{Journal of Physics: Materials}, 2019.
  https://doi.org/10.1088/2515-7639/ab084b\relax
\mciteBstWouldAddEndPuncttrue
\mciteSetBstMidEndSepPunct{\mcitedefaultmidpunct}
{\mcitedefaultendpunct}{\mcitedefaultseppunct}\relax
\EndOfBibitem
\bibitem[Kresse and Furthm{\"{u}}ller(1996)]{vasp1}
G.~Kresse and J.~Furthm{\"{u}}ller, \emph{Comput. Mater. Sci.}, 1996,
  \textbf{6}, 15--50\relax
\mciteBstWouldAddEndPuncttrue
\mciteSetBstMidEndSepPunct{\mcitedefaultmidpunct}
{\mcitedefaultendpunct}{\mcitedefaultseppunct}\relax
\EndOfBibitem
\bibitem[Kresse and Furthmuller(1996)]{vasp2}
G.~Kresse and J.~Furthmuller, \emph{Phys. Rev. B}, 1996, \textbf{54},
  11169--11186\relax
\mciteBstWouldAddEndPuncttrue
\mciteSetBstMidEndSepPunct{\mcitedefaultmidpunct}
{\mcitedefaultendpunct}{\mcitedefaultseppunct}\relax
\EndOfBibitem
\bibitem[Landauer(1970)]{landauer1}
R.~Landauer, \emph{The Philosophical Magazine: A Journal of Theoretical
  Experimental and Applied Physics}, 1970, \textbf{21}, 863--867\relax
\mciteBstWouldAddEndPuncttrue
\mciteSetBstMidEndSepPunct{\mcitedefaultmidpunct}
{\mcitedefaultendpunct}{\mcitedefaultseppunct}\relax
\EndOfBibitem
\bibitem[Novaes \emph{et~al.}(2006)Novaes, Silva, and Fazzio]{NOVAES2006}
F.~D. Novaes, A.~A. J. R.~d. Silva and A.~Fazzio, \emph{{Brazilian Journal of
  Physics}}, 2006, \textbf{36}, 799 -- 807\relax
\mciteBstWouldAddEndPuncttrue
\mciteSetBstMidEndSepPunct{\mcitedefaultmidpunct}
{\mcitedefaultendpunct}{\mcitedefaultseppunct}\relax
\EndOfBibitem
\bibitem[Brandbyge \emph{et~al.}(2002)Brandbyge, Mozos, Ordej\'on, Taylor, and
  Stokbro]{PhysRevB.65.165401}
M.~Brandbyge, J.-L. Mozos, P.~Ordej\'on, J.~Taylor and K.~Stokbro, \emph{Phys.
  Rev. B}, 2002, \textbf{65}, 165401\relax
\mciteBstWouldAddEndPuncttrue
\mciteSetBstMidEndSepPunct{\mcitedefaultmidpunct}
{\mcitedefaultendpunct}{\mcitedefaultseppunct}\relax
\EndOfBibitem
\bibitem[Lima \emph{et~al.}(2011)Lima, da~Silva, and
  Fazzio]{PhysRevB.84.245411}
M.~P. Lima, A.~J.~R. da~Silva and A.~Fazzio, \emph{Phys. Rev. B}, 2011,
  \textbf{84}, 245411\relax
\mciteBstWouldAddEndPuncttrue
\mciteSetBstMidEndSepPunct{\mcitedefaultmidpunct}
{\mcitedefaultendpunct}{\mcitedefaultseppunct}\relax
\EndOfBibitem
\bibitem[{Lima} \emph{et~al.}(2018){Lima}, {Fazzio}, and {da Silva}]{8387867}
M.~P. {Lima}, A.~{Fazzio} and A.~J.~R. {da Silva}, \emph{IEEE Electron Device
  Letters}, 2018, \textbf{39}, 1258--1261\relax
\mciteBstWouldAddEndPuncttrue
\mciteSetBstMidEndSepPunct{\mcitedefaultmidpunct}
{\mcitedefaultendpunct}{\mcitedefaultseppunct}\relax
\EndOfBibitem
\bibitem[Rocha \emph{et~al.}(2008)Rocha, Rossi, Fazzio, and
  da~Silva]{recursive1}
A.~R. Rocha, M.~Rossi, A.~Fazzio and A.~J.~R. da~Silva, \emph{Phys. Rev.
  Lett.}, 2008, \textbf{100}, 176803\relax
\mciteBstWouldAddEndPuncttrue
\mciteSetBstMidEndSepPunct{\mcitedefaultmidpunct}
{\mcitedefaultendpunct}{\mcitedefaultseppunct}\relax
\EndOfBibitem
\bibitem[Rocha \emph{et~al.}(2010)Rocha, Rossi, da~Silva, and
  Fazzio]{recursive2}
A.~R. Rocha, M.~Rossi, A.~J.~R. da~Silva and A.~Fazzio, \emph{Journal of
  Physics D: Applied Physics}, 2010, \textbf{43}, 374002\relax
\mciteBstWouldAddEndPuncttrue
\mciteSetBstMidEndSepPunct{\mcitedefaultmidpunct}
{\mcitedefaultendpunct}{\mcitedefaultseppunct}\relax
\EndOfBibitem
\bibitem[Mostofi \emph{et~al.}(2014)Mostofi, Yates, Pizzi, Lee, Souza,
  Vanderbilt, and Marzari]{MOSTOFI20142309}
A.~A. Mostofi, J.~R. Yates, G.~Pizzi, Y.-S. Lee, I.~Souza, D.~Vanderbilt and
  N.~Marzari, \emph{Computer Physics Communications}, 2014, \textbf{185}, 2309
  -- 2310\relax
\mciteBstWouldAddEndPuncttrue
\mciteSetBstMidEndSepPunct{\mcitedefaultmidpunct}
{\mcitedefaultendpunct}{\mcitedefaultseppunct}\relax
\EndOfBibitem
\bibitem[Leenaerts \emph{et~al.}(2008)Leenaerts, Partoens, and
  Peeters]{PhysRevB.77.125416}
O.~Leenaerts, B.~Partoens and F.~M. Peeters, \emph{Phys. Rev. B}, 2008,
  \textbf{77}, 125416\relax
\mciteBstWouldAddEndPuncttrue
\mciteSetBstMidEndSepPunct{\mcitedefaultmidpunct}
{\mcitedefaultendpunct}{\mcitedefaultseppunct}\relax
\EndOfBibitem
\bibitem[Sangwan and Hersam(2018)]{doi:10.1146/annurev-physchem-050317-021353}
V.~K. Sangwan and M.~C. Hersam, \emph{Annual Review of Physical Chemistry},
  2018, \textbf{69}, 299--325\relax
\mciteBstWouldAddEndPuncttrue
\mciteSetBstMidEndSepPunct{\mcitedefaultmidpunct}
{\mcitedefaultendpunct}{\mcitedefaultseppunct}\relax
\EndOfBibitem
\bibitem[Bhandari \emph{et~al.}(2017)Bhandari, Wang, Watanabe, Taniguchi, Kim,
  and Westervelt]{Bhandari_2017}
S.~Bhandari, K.~Wang, K.~Watanabe, T.~Taniguchi, P.~Kim and R.~M. Westervelt,
  \emph{Journal of Physics: Conference Series}, 2017, \textbf{864},
  012031\relax
\mciteBstWouldAddEndPuncttrue
\mciteSetBstMidEndSepPunct{\mcitedefaultmidpunct}
{\mcitedefaultendpunct}{\mcitedefaultseppunct}\relax
\EndOfBibitem
\bibitem[Ghatak \emph{et~al.}(2011)Ghatak, Pal, and
  Ghosh]{doi:10.1021/nn202852j}
S.~Ghatak, A.~N. Pal and A.~Ghosh, \emph{ACS Nano}, 2011, \textbf{5},
  7707--7712\relax
\mciteBstWouldAddEndPuncttrue
\mciteSetBstMidEndSepPunct{\mcitedefaultmidpunct}
{\mcitedefaultendpunct}{\mcitedefaultseppunct}\relax
\EndOfBibitem
\bibitem[Biel \emph{et~al.}(2009)Biel, Blase, Triozon, and
  Roche]{PhysRevLett.102.096803}
B.~Biel, X.~Blase, F.~m.~c. Triozon and S.~Roche, \emph{Phys. Rev. Lett.},
  2009, \textbf{102}, 096803\relax
\mciteBstWouldAddEndPuncttrue
\mciteSetBstMidEndSepPunct{\mcitedefaultmidpunct}
{\mcitedefaultendpunct}{\mcitedefaultseppunct}\relax
\EndOfBibitem
\bibitem[Deretzis \emph{et~al.}(2010)Deretzis, Fiori, Iannaccone, and
  La~Magna]{PhysRevB.81.085427}
I.~Deretzis, G.~Fiori, G.~Iannaccone and A.~La~Magna, \emph{Phys. Rev. B},
  2010, \textbf{81}, 085427\relax
\mciteBstWouldAddEndPuncttrue
\mciteSetBstMidEndSepPunct{\mcitedefaultmidpunct}
{\mcitedefaultendpunct}{\mcitedefaultseppunct}\relax
\EndOfBibitem
\bibitem[López-Bezanilla \emph{et~al.}(2009)López-Bezanilla, Triozon, Latil,
  Blase, and Roche]{doi:10.1021/nl802798q}
A.~López-Bezanilla, F.~Triozon, S.~Latil, X.~Blase and S.~Roche, \emph{Nano
  Letters}, 2009, \textbf{9}, 940--944\relax
\mciteBstWouldAddEndPuncttrue
\mciteSetBstMidEndSepPunct{\mcitedefaultmidpunct}
{\mcitedefaultendpunct}{\mcitedefaultseppunct}\relax
\EndOfBibitem
\bibitem[Baugher \emph{et~al.}(2013)Baugher, Churchill, Yang, and
  Jarillo-Herrero]{doi:10.1021/nl401916s}
B.~W.~H. Baugher, H.~O.~H. Churchill, Y.~Yang and P.~Jarillo-Herrero,
  \emph{Nano Letters}, 2013, \textbf{13}, 4212--4216\relax
\mciteBstWouldAddEndPuncttrue
\mciteSetBstMidEndSepPunct{\mcitedefaultmidpunct}
{\mcitedefaultendpunct}{\mcitedefaultseppunct}\relax
\EndOfBibitem
\bibitem[Schmidt \emph{et~al.}(2014)Schmidt, Wang, Chu, Toh, Kumar, Zhao,
  Castro~Neto, Martin, Adam, Özyilmaz, and Eda]{doi:10.1021/nl4046922}
H.~Schmidt, S.~Wang, L.~Chu, M.~Toh, R.~Kumar, W.~Zhao, A.~H. Castro~Neto,
  J.~Martin, S.~Adam, B.~Özyilmaz and G.~Eda, \emph{Nano Letters}, 2014,
  \textbf{14}, 1909--1913\relax
\mciteBstWouldAddEndPuncttrue
\mciteSetBstMidEndSepPunct{\mcitedefaultmidpunct}
{\mcitedefaultendpunct}{\mcitedefaultseppunct}\relax
\EndOfBibitem
\bibitem[Jariwala \emph{et~al.}(2013)Jariwala, Sangwan, Late, Johns, Dravid,
  Marks, Lauhon, and Hersam]{doi:10.1063/1.4803920}
D.~Jariwala, V.~K. Sangwan, D.~J. Late, J.~E. Johns, V.~P. Dravid, T.~J. Marks,
  L.~J. Lauhon and M.~C. Hersam, \emph{Applied Physics Letters}, 2013,
  \textbf{102}, 173107\relax
\mciteBstWouldAddEndPuncttrue
\mciteSetBstMidEndSepPunct{\mcitedefaultmidpunct}
{\mcitedefaultendpunct}{\mcitedefaultseppunct}\relax
\EndOfBibitem
\end{mcitethebibliography}


\providecommand{\latin}[1]{#1}
\providecommand*\mcitethebibliography{\thebibliography}
\csname @ifundefined\endcsname{endmcitethebibliography}
  {\let\endmcitethebibliography\endthebibliography}{}
\begin{mcitethebibliography}{24}
\providecommand*\natexlab[1]{#1}
\providecommand*\mciteSetBstSublistMode[1]{}
\providecommand*\mciteSetBstMaxWidthForm[2]{}
\providecommand*\mciteBstWouldAddEndPuncttrue
  {\def\EndOfBibitem{\unskip.}}
\providecommand*\mciteBstWouldAddEndPunctfalse
  {\let\EndOfBibitem\relax}
\providecommand*\mciteSetBstMidEndSepPunct[3]{}
\providecommand*\mciteSetBstSublistLabelBeginEnd[3]{}
\providecommand*\EndOfBibitem{}
\mciteSetBstSublistMode{f}
\mciteSetBstMaxWidthForm{subitem}{(\alph{mcitesubitemcount})}
\mciteSetBstSublistLabelBeginEnd
  {\mcitemaxwidthsubitemform\space}
  {\relax}
  {\relax}

\bibitem[Novoselov \latin{et~al.}(2004)Novoselov, Geim, Morozov, Jiang, Zhang,
  Dubonos, Grigorieva, and Firsov]{Novoselov666}
Novoselov,~K.~S.; Geim,~A.~K.; Morozov,~S.~V.; Jiang,~D.; Zhang,~Y.;
  Dubonos,~S.~V.; Grigorieva,~I.~V.; Firsov,~A.~A. Electric Field Effect in
  Atomically Thin Carbon Films. \emph{Science} \textbf{2004}, \emph{306},
  666--669\relax
\mciteBstWouldAddEndPuncttrue
\mciteSetBstMidEndSepPunct{\mcitedefaultmidpunct}
{\mcitedefaultendpunct}{\mcitedefaultseppunct}\relax
\EndOfBibitem
\bibitem[Castro~Neto \latin{et~al.}(2009)Castro~Neto, Guinea, Peres, Novoselov,
  and Geim]{RevModPhys.81.109}
Castro~Neto,~A.~H.; Guinea,~F.; Peres,~N. M.~R.; Novoselov,~K.~S.; Geim,~A.~K.
  The electronic properties of graphene. \emph{Rev. Mod. Phys.} \textbf{2009},
  \emph{81}, 109--162\relax
\mciteBstWouldAddEndPuncttrue
\mciteSetBstMidEndSepPunct{\mcitedefaultmidpunct}
{\mcitedefaultendpunct}{\mcitedefaultseppunct}\relax
\EndOfBibitem
\bibitem[Han \latin{et~al.}(2015)Han, Bhatia, and Kim]{Han2015}
Han,~S.~A.; Bhatia,~R.; Kim,~S.-W. Synthesis, properties and potential
  applications of two-dimensional transition metal dichalcogenides. \emph{Nano
  Convergence} \textbf{2015}, \emph{2}, 17\relax
\mciteBstWouldAddEndPuncttrue
\mciteSetBstMidEndSepPunct{\mcitedefaultmidpunct}
{\mcitedefaultendpunct}{\mcitedefaultseppunct}\relax
\EndOfBibitem
\bibitem[Korm\'{a}nyos \latin{et~al.}(2015)Korm\'{a}nyos, Burkard, Gmitra,
  Fabian, Z\'{o}lyomi, Drummond, and Fal'ko]{2053-1583-2-2-022001}
Korm\'{a}nyos,~A.; Burkard,~G.; Gmitra,~M.; Fabian,~J.; Z\'{o}lyomi,~V.;
  Drummond,~N.~D.; Fal'ko,~V. k.p theory for two-dimensional transition metal
  dichalcogenide semiconductors. \emph{2D Materials} \textbf{2015}, \emph{2},
  022001\relax
\mciteBstWouldAddEndPuncttrue
\mciteSetBstMidEndSepPunct{\mcitedefaultmidpunct}
{\mcitedefaultendpunct}{\mcitedefaultseppunct}\relax
\EndOfBibitem
\bibitem[Pan and Zhang(2012)Pan, and Zhang]{C2JM15906F}
Pan,~H.; Zhang,~Y.-W. Edge-dependent structural{,} electronic and magnetic
  properties of MoS2 nanoribbons. \emph{J. Mater. Chem.} \textbf{2012},
  \emph{22}, 7280--7290\relax
\mciteBstWouldAddEndPuncttrue
\mciteSetBstMidEndSepPunct{\mcitedefaultmidpunct}
{\mcitedefaultendpunct}{\mcitedefaultseppunct}\relax
\EndOfBibitem
\bibitem[Qian \latin{et~al.}(2014)Qian, Liu, Fu, and Li]{Qian1344}
Qian,~X.; Liu,~J.; Fu,~L.; Li,~J. Quantum spin Hall effect in two-dimensional
  transition metal dichalcogenides. \emph{Science} \textbf{2014}, \emph{346},
  1344--1347\relax
\mciteBstWouldAddEndPuncttrue
\mciteSetBstMidEndSepPunct{\mcitedefaultmidpunct}
{\mcitedefaultendpunct}{\mcitedefaultseppunct}\relax
\EndOfBibitem
\bibitem[An \latin{et~al.}(2016)An, Zhang, Da, Fu, Jiao, and
  Liu]{0022-3727-49-24-245304}
An,~Y.; Zhang,~M.; Da,~H.; Fu,~Z.; Jiao,~Z.; Liu,~Z. Width and defect effects
  on the electronic transport of zigzag MoS 2 nanoribbons. \emph{Journal of
  Physics D: Applied Physics} \textbf{2016}, \emph{49}, 245304\relax
\mciteBstWouldAddEndPuncttrue
\mciteSetBstMidEndSepPunct{\mcitedefaultmidpunct}
{\mcitedefaultendpunct}{\mcitedefaultseppunct}\relax
\EndOfBibitem
\bibitem[Coleman \latin{et~al.}(2011)Coleman, Lotya, O{\textquoteright}Neill,
  Bergin, King, Khan, Young, Gaucher, De, Smith, Shvets, Arora, Stanton, Kim,
  Lee, Kim, Duesberg, Hallam, Boland, Wang, Donegan, Grunlan, Moriarty,
  Shmeliov, Nicholls, Perkins, Grieveson, Theuwissen, McComb, Nellist, and
  Nicolosi]{chem1}
Coleman,~J.~N. \latin{et~al.}  Two-Dimensional Nanosheets Produced by Liquid
  Exfoliation of Layered Materials. \emph{Science} \textbf{2011}, \emph{331},
  568--571\relax
\mciteBstWouldAddEndPuncttrue
\mciteSetBstMidEndSepPunct{\mcitedefaultmidpunct}
{\mcitedefaultendpunct}{\mcitedefaultseppunct}\relax
\EndOfBibitem
\bibitem[Chhowalla \latin{et~al.}(2013)Chhowalla, Shin, Eda, Li, Loh, and
  Zhang]{chem2}
Chhowalla,~M.; Shin,~H.~S.; Eda,~G.; Li,~L.-J.; Loh,~K.~P.; Zhang,~H. The
  chemistry of two-dimensional layered transition metal dichalcogenide
  nanosheets. \emph{Nature Chemistry} \textbf{2013}, \emph{5}, 263 EP --,
  Review Article\relax
\mciteBstWouldAddEndPuncttrue
\mciteSetBstMidEndSepPunct{\mcitedefaultmidpunct}
{\mcitedefaultendpunct}{\mcitedefaultseppunct}\relax
\EndOfBibitem
\bibitem[Gon{\c{c}}alves \latin{et~al.}(2015)Gon{\c{c}}alves, Fiel, Soares,
  Schreiner, Silva, and Leite]{doi:10.1002/chem.201502303}
Gon{\c{c}}alves,~R.~H.; Fiel,~R.; Soares,~M. R.~S.; Schreiner,~W.~H.; Silva,~C.
  M.~P.; Leite,~E.~R. Single-Step Exfoliation and Covalent Functionalization of
  MoS2 Nanosheets by an Organosulfur Reaction. \emph{Chemistry - A European
  Journal} \textbf{2015}, \emph{21}, 15583--15588\relax
\mciteBstWouldAddEndPuncttrue
\mciteSetBstMidEndSepPunct{\mcitedefaultmidpunct}
{\mcitedefaultendpunct}{\mcitedefaultseppunct}\relax
\EndOfBibitem
\bibitem[Rocha \latin{et~al.}(2008)Rocha, Rossi, Fazzio, and
  da~Silva]{recursive1}
Rocha,~A.~R.; Rossi,~M.; Fazzio,~A.; da~Silva,~A. J.~R. Designing Real
  Nanotube-Based Gas Sensors. \emph{Phys. Rev. Lett.} \textbf{2008},
  \emph{100}, 176803\relax
\mciteBstWouldAddEndPuncttrue
\mciteSetBstMidEndSepPunct{\mcitedefaultmidpunct}
{\mcitedefaultendpunct}{\mcitedefaultseppunct}\relax
\EndOfBibitem
\bibitem[Hohenberg and Kohn(1964)Hohenberg, and Kohn]{DFT1}
Hohenberg,~P.; Kohn,~W. {Inhomogeneous Electron Gas}. \emph{Physical Review}
  \textbf{1964}, \emph{136}, B864--B871\relax
\mciteBstWouldAddEndPuncttrue
\mciteSetBstMidEndSepPunct{\mcitedefaultmidpunct}
{\mcitedefaultendpunct}{\mcitedefaultseppunct}\relax
\EndOfBibitem
\bibitem[Kresse and Furthm{\"{u}}ller(1996)Kresse, and
  Furthm{\"{u}}ller]{vasp1}
Kresse,~G.; Furthm{\"{u}}ller,~J. {Efficiency of ab-initio total energy
  calculations for metals and semiconductors using a plane-wave basis set}.
  \emph{Comput. Mater. Sci.} \textbf{1996}, \emph{6}, 15--50\relax
\mciteBstWouldAddEndPuncttrue
\mciteSetBstMidEndSepPunct{\mcitedefaultmidpunct}
{\mcitedefaultendpunct}{\mcitedefaultseppunct}\relax
\EndOfBibitem
\bibitem[Kresse and Furthmuller(1996)Kresse, and Furthmuller]{vasp2}
Kresse,~G.; Furthmuller,~J. {Efficient iterative schemes for ab initio
  total-energy calculations using a plane-wave basis set}. \emph{Phys. Rev. B}
  \textbf{1996}, \emph{54}, 11169--11186\relax
\mciteBstWouldAddEndPuncttrue
\mciteSetBstMidEndSepPunct{\mcitedefaultmidpunct}
{\mcitedefaultendpunct}{\mcitedefaultseppunct}\relax
\EndOfBibitem
\bibitem[Kohn and Sham(1965)Kohn, and Sham]{DFT2}
Kohn,~W.; Sham,~L.~J. {Self-Consistent Equations Including Exchange and
  Correlation Effects}. \emph{Physical Review} \textbf{1965}, \emph{140},
  A1133--A1138\relax
\mciteBstWouldAddEndPuncttrue
\mciteSetBstMidEndSepPunct{\mcitedefaultmidpunct}
{\mcitedefaultendpunct}{\mcitedefaultseppunct}\relax
\EndOfBibitem
\bibitem[Soler \latin{et~al.}(2002)Soler, Artacho, Gale, García, Junquera,
  Ordejón, and Sánchez-Portal]{0953-8984-14-11-302}
Soler,~J.~M.; Artacho,~E.; Gale,~J.~D.; García,~A.; Junquera,~J.;
  Ordejón,~P.; Sánchez-Portal,~D. The SIESTA method for ab initio order- N
  materials simulation. \emph{Journal of Physics: Condensed Matter}
  \textbf{2002}, \emph{14}, 2745\relax
\mciteBstWouldAddEndPuncttrue
\mciteSetBstMidEndSepPunct{\mcitedefaultmidpunct}
{\mcitedefaultendpunct}{\mcitedefaultseppunct}\relax
\EndOfBibitem
\bibitem[Rocha \latin{et~al.}(2010)Rocha, Rossi, da~Silva, and
  Fazzio]{recursive2}
Rocha,~A.~R.; Rossi,~M.; da~Silva,~A. J.~R.; Fazzio,~A. Realistic calculations
  of carbon-based disordered systems. \emph{Journal of Physics D: Applied
  Physics} \textbf{2010}, \emph{43}, 374002\relax
\mciteBstWouldAddEndPuncttrue
\mciteSetBstMidEndSepPunct{\mcitedefaultmidpunct}
{\mcitedefaultendpunct}{\mcitedefaultseppunct}\relax
\EndOfBibitem
\bibitem[Biel \latin{et~al.}(2009)Biel, Blase, Triozon, and
  Roche]{PhysRevLett.102.096803}
Biel,~B.; Blase,~X.; Triozon,~F. m.~c.; Roche,~S. Anomalous Doping Effects on
  Charge Transport in Graphene Nanoribbons. \emph{Phys. Rev. Lett.}
  \textbf{2009}, \emph{102}, 096803\relax
\mciteBstWouldAddEndPuncttrue
\mciteSetBstMidEndSepPunct{\mcitedefaultmidpunct}
{\mcitedefaultendpunct}{\mcitedefaultseppunct}\relax
\EndOfBibitem
\bibitem[Deretzis \latin{et~al.}(2010)Deretzis, Fiori, Iannaccone, and
  La~Magna]{PhysRevB.81.085427}
Deretzis,~I.; Fiori,~G.; Iannaccone,~G.; La~Magna,~A. Effects due to
  backscattering and pseudogap features in graphene nanoribbons with single
  vacancies. \emph{Phys. Rev. B} \textbf{2010}, \emph{81}, 085427\relax
\mciteBstWouldAddEndPuncttrue
\mciteSetBstMidEndSepPunct{\mcitedefaultmidpunct}
{\mcitedefaultendpunct}{\mcitedefaultseppunct}\relax
\EndOfBibitem
\bibitem[López-Bezanilla \latin{et~al.}(2009)López-Bezanilla, Triozon, Latil,
  Blase, and Roche]{doi:10.1021/nl802798q}
López-Bezanilla,~A.; Triozon,~F.; Latil,~S.; Blase,~X.; Roche,~S. Effect of
  the Chemical Functionalization on Charge Transport in Carbon Nanotubes at the
  Mesoscopic Scale. \emph{Nano Letters} \textbf{2009}, \emph{9}, 940--944,
  PMID: 19191494\relax
\mciteBstWouldAddEndPuncttrue
\mciteSetBstMidEndSepPunct{\mcitedefaultmidpunct}
{\mcitedefaultendpunct}{\mcitedefaultseppunct}\relax
\EndOfBibitem
\bibitem[Baugher \latin{et~al.}(2013)Baugher, Churchill, Yang, and
  Jarillo-Herrero]{doi:10.1021/nl401916s}
Baugher,~B. W.~H.; Churchill,~H. O.~H.; Yang,~Y.; Jarillo-Herrero,~P. Intrinsic
  Electronic Transport Properties of High-Quality Monolayer and Bilayer MoS2.
  \emph{Nano Letters} \textbf{2013}, \emph{13}, 4212--4216, PMID:
  23930826\relax
\mciteBstWouldAddEndPuncttrue
\mciteSetBstMidEndSepPunct{\mcitedefaultmidpunct}
{\mcitedefaultendpunct}{\mcitedefaultseppunct}\relax
\EndOfBibitem
\bibitem[Schmidt \latin{et~al.}(2014)Schmidt, Wang, Chu, Toh, Kumar, Zhao,
  Castro~Neto, Martin, Adam, Özyilmaz, and Eda]{doi:10.1021/nl4046922}
Schmidt,~H.; Wang,~S.; Chu,~L.; Toh,~M.; Kumar,~R.; Zhao,~W.;
  Castro~Neto,~A.~H.; Martin,~J.; Adam,~S.; Özyilmaz,~B.; Eda,~G. Transport
  Properties of Monolayer MoS2 Grown by Chemical Vapor Deposition. \emph{Nano
  Letters} \textbf{2014}, \emph{14}, 1909--1913, PMID: 24640984\relax
\mciteBstWouldAddEndPuncttrue
\mciteSetBstMidEndSepPunct{\mcitedefaultmidpunct}
{\mcitedefaultendpunct}{\mcitedefaultseppunct}\relax
\EndOfBibitem
\bibitem[Jariwala \latin{et~al.}(2013)Jariwala, Sangwan, Late, Johns, Dravid,
  Marks, Lauhon, and Hersam]{doi:10.1063/1.4803920}
Jariwala,~D.; Sangwan,~V.~K.; Late,~D.~J.; Johns,~J.~E.; Dravid,~V.~P.;
  Marks,~T.~J.; Lauhon,~L.~J.; Hersam,~M.~C. Band-like transport in high
  mobility unencapsulated single-layer MoS2 transistors. \emph{Applied Physics
  Letters} \textbf{2013}, \emph{102}, 173107\relax
\mciteBstWouldAddEndPuncttrue
\mciteSetBstMidEndSepPunct{\mcitedefaultmidpunct}
{\mcitedefaultendpunct}{\mcitedefaultseppunct}\relax
\EndOfBibitem
\end{mcitethebibliography}
\bibliographystyle{rsc} 

\end{document}